\documentclass[11pt,a4paper,final]{article}
\usepackage[left=1in,right=1in,top=1in,bottom=1in]{geometry}
\usepackage{authblk}
\PassOptionsToPackage{sort&compress}{natbib}

\usepackage[numbers,super]{natbib}
\usepackage{lineno}

\usepackage{afterpage}
\usepackage{floatrow}
\usepackage{booktabs} % For formal tables
\usepackage[ruled]{algorithm2e} % For algorithms
\usepackage{tikz}
\usetikzlibrary{arrows}
\usetikzlibrary{shapes.multipart}
\usepackage{caption}

\SetAlFnt{\small}
\SetAlCapFnt{\small}
\SetAlCapNameFnt{\small}
\SetAlCapHSkip{0pt}
\IncMargin{-\parindent}
\usepackage[all]{nowidow}
\usepackage{lscape} 
\usepackage{float}

\usepackage[font=small,labelfont=bf]{caption}
\usepackage[utf8]{inputenc} % allow utf-8 input
\usepackage[T1]{fontenc}    % use 8-bit T1 fonts
\usepackage{url}            % simple URL typesetting
\usepackage{hyperref}       % hyperlinks
\hypersetup{
    colorlinks=true,
    linkcolor=blue,
    filecolor=magenta,      
    urlcolor=cyan,
    citecolor=gray,
    pdfpagemode=FullScreen,
}
\usepackage{booktabs}       % professional-quality tables
\usepackage{amsfonts}       % blackboard math symbols
\usepackage{nicefrac}       % compact symbols for 1/2, etc.
\usepackage{microtype}      % microtypography
\usepackage{xcolor}         % colors
\usepackage{amsmath}
\usepackage{mathtools}
\usepackage{amsthm}
\usepackage{amsmath}
\usepackage{amssymb}
\usepackage{bbm}
\usepackage{tikz} 
\usepackage{wrapfig}
\usepackage{floatrow}
\usepackage{enumitem}

\usepackage[suppress]{color-edits}

\usepackage{caption}
\usepackage{subcaption}
\usepackage{rotating}

\title{Building Machines that Learn and Think \emph{with} People}

\begin{document}
\nolinenumbers
\author[1]{Katherine M. Collins\footnote{Contributed equally.}}
\author[2]{Ilia Sucholutsky$^*$}
\author[3,4]{Umang Bhatt$^*$}
\author[5]{Kartik Chandra$^*$}
\author[5]{Lionel Wong$^*$}
\author[6,7]{Mina Lee\footnote{Contributed equally.}}
\author[5]{Cedegao E. Zhang$^\dagger$}
\author[5]{Tan Zhi-Xuan$^\dagger$}
\author[3]{Mark Ho$^\dagger$}
\author[5]{Vikash Mansinghka\footnote{Equal senior role.}}
\author[1,4]{Adrian Weller$^\ddagger$}
\author[5]{Joshua B. Tenenbaum$^\ddagger$}
\author[2]{Thomas L. Griffiths$^\ddagger$}
\affil[1]{University of Cambridge}
\affil[2]{Princeton University}
\affil[3]{NYU}
\affil[4]{The Alan Turing Institute}
\affil[5]{MIT}
\affil[6]{Microsoft Research}
\affil[7]{University of Chicago}
\date{}

\maketitle
%TC:ignore

% \tableofcontents
% \clearpage

\begin{abstract}

What do we want from machine intelligence? We envision machines that are not just \emph{tools} for thought, but \textit{partners} in thought: reasonable, insightful, knowledgeable, reliable, and trustworthy systems that think \emph{with} us. Current artificial intelligence (AI) systems satisfy some of these criteria, some of the time. In this Perspective, we show how the science of collaborative cognition can be put to work to engineer systems that really can be called ``thought partners,'' systems built to meet our expectations and complement our limitations. We lay out several modes of collaborative thought in which humans and AI thought partners can engage and propose desiderata for human-compatible thought partnerships. Drawing on motifs from computational cognitive science, we motivate an alternative scaling path for the design of thought partners and ecosystems around their use through a Bayesian lens, whereby the partners we construct actively build and reason over models of the human and world.

\end{abstract}

\section{Introduction}

Computers have long been seen as tools for thought. Steve Jobs called computers ``bicycles for the mind'': tools that dramatically increase the efficiency, productivity, and joy of thinking. Now, thirty years later, this metaphor is beginning to change. Computer systems are increasingly referred to not as \emph{vehicles} but as ``copilots''\cite{GitHubCopilot, CopilotMicrosoft}: we have moved from designing \textit{tools} for thought to actual \textit{partners} in thought. 

The current wave of AI technologies, particularly language models, have catalyzed this transition. Users no longer have to know how to write code to engage intimately with computers; we can now interface through the medium of natural language. Humans already think alone and together, at least communicated often through the medium of language~\cite{fedorenko2024language}. We long have -- from developing new modes of thinking through questioning and debate to teaching and learning through language. The apparent power of these new systems -- getting closer to the kind of artificial intelligence (AI) imagined in the field's early days~\citep{turing1950, clynes1960cyborgs, weizenbaum1966eliza, benHCAI, Bundy1983TheCM, anderson1990cognitive} -- as well as challenges faced by the current iterations of such systems -- invites us to think about what it will take to build systems that truly act as effective thought partners. We argue that good thought partners are systems (1) which can understand us, (2) which we can understand, and (3) which have sufficient understanding of the world that we can engage on common ground.

One path to building such thought partners is to scale foundation models (e.g., LLMs~\citep{foundationModels}) with large amounts of human demonstrations and feedback, along with ``traces'' of human thought scraped from web-scale data~\citep{ouyang2022training, christiano2017deep, lee2023aligning}. While such an approach has produced systems that accurately mimic human \textit{behavior} (e.g., producing fluent text), these machines do not robustly simulate human \textit{cognition} (e.g., explicitly reasoning about the world or other minds) in ways expected by a true thought partner~\citep{ullman2023large, collins2024evaluating, wong2023word, zhang2023ai, gweon2023socially, mahowald2024dissociating, fedorenko2024language, mccoy2023embers}. 

What would it take to design systems that meet our criteria? One promising path is to design systems that build explicit models of the task, world, and human (where these models are structured~\citep{tenenbaum2011grow}, rather than distributionally learned from data) -- drawing on formal frameworks grounded in cognitive psychology for understanding how humans think, alone and together. In this Perspective, we chart a new vision for the design of AI thought partners. Decades of work in the behavioral sciences provide valuable insights for designing human-centric, uncertainty-aware thought partners. Drawing on such research, we argue that effective thought partners are those which \textit{build models of the human} and \textit{the world}. 

This toolkit includes foundation models~\citep{ griffiths2023bayes, sumers2023cognitive, binz2023turning}, but is not limited to them. Indeed, foundation models like LLMs are fueling new motifs for thinking about human minds in computational terms (e.g., ``rational meaning construction''~\citep{wong2023word}) interleaved \textit{alongside} techniques from probabilistic programming~\citep{cusumano2019gen, goodman2008church, bingham2019pyro, ge2018turing, goodman2014concepts}, goal-directed search~\citep{van2023expertise, trinh2024solving, yao2024tree}, and other explicit, structured representations, e.g., of agents thinking about other agents~\citep{baker2009action,jara2020naive, zhi2024pragmatic}. We already have tools that help us build machines that learn and think \textit{like} people~\citep{lake_ullman_tenenbaum_gershman_2017}. We propose applying that toolkit to collaborative cognition -- to build machines that learn and think \textit{with} people.

\begin{figure}
    \centering
    \includegraphics[width=0.8\textwidth]{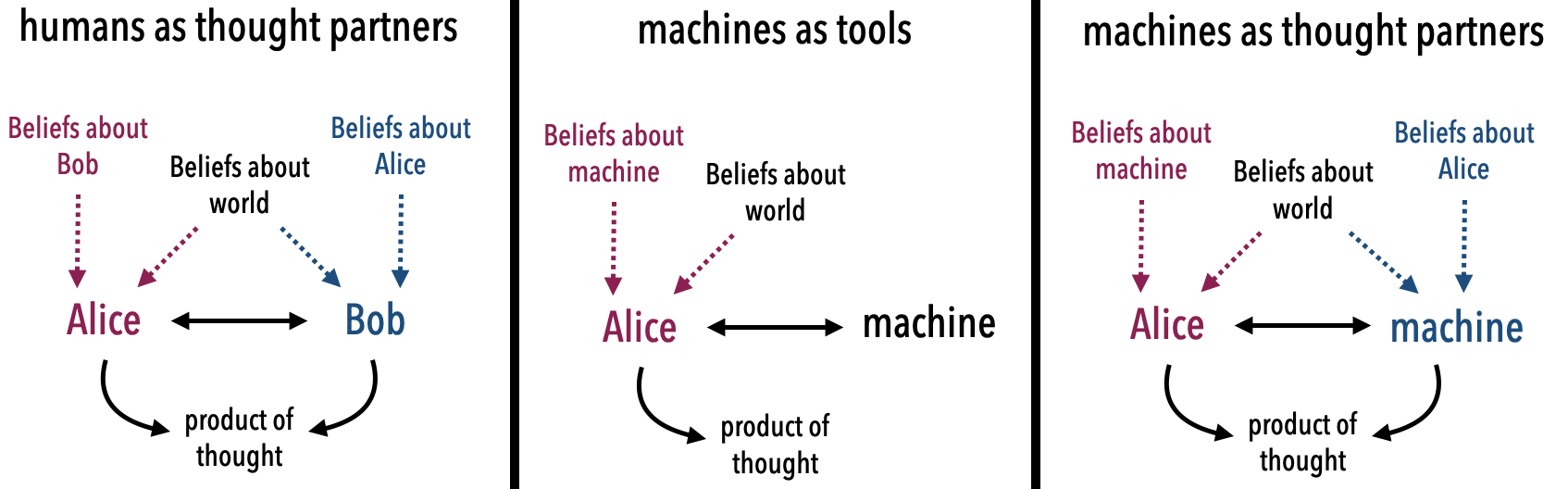}

\caption{\textbf{Examples of ecosystems for thinking.} Humans have long thought together. Machines expanded the efficiency of human thinking. Now, machines -- powered by AI -- open up new realms of computational thought partnership with humans.}
    \label{fig:framework}
\end{figure}

\section{What are Thought Partners?}
\label{what-thought-partners}

When we think, we draw coherent inferences, make predictions, and act on these predictions -- from assessing what birthday present to gift a treasured friend, to formulating a new scientific hypothesis and experiment plan to evaluate a theory. We flexibly draw on prior knowledge and update our beliefs through experience (as we discuss below). We not only solve problems, but imagine new ones~\citep{chu2020play}. And we \textit{think together}. For generations, humans have discussed and debated ideas, and developed ecosystems to disseminate such thoughts to new audiences. Much scientific innovation has come through collaboration, where advances are frequently fueled by engaging with diverse partners who offer new ideas yet share our values~\citep{yanai2024takes}.

\subsection{Modes of Collaborative Thought}
\label{modes-of-thought-txt}

As an illustration of the many ways that people and machines might think with each other, 
we highlight a few \textit{modes of collaborative thought} (Table \ref{tab:modes_of_thought}). This set of modes, partly inspired by characterizations of thinking and reasoning in psychology \citep{holyoak2005cambridge, holyoak2012oxford}, are not meant to be comprehensive of all aspects of thought. Rather, we see these modes as ripe for the further development of AI thought partners. 

% \subsection{Case Studies}
\subsection{Example Domains}
\label{case-study-intro}

We next outline a few diverse domains in which the development of AI thought partners able to truly collaborate with humans (Figure~\ref{fig:framework}) may be particularly valuable. We highlight common computational challenges that arise when considering what effective partnership might look like in each domain, foreshadowing our proposed desiderata. We later return to these case studies with concrete human-centric thought partner instantiations. 

\textit{Thought Partners for Programming.} Programming is a cognitively-demanding activity that requires gaining fluency in translating human intentions into formal, machine-interpretable languages. It is no surprise that decades of effort have gone into designing tools to help people program~\citep{ko2004designing, ko2011state, muggleton1994inductive, anderson1985lisp, anderson1995cognitive}. New ``programming assistant'' tools like GitHub Copilot have rapidly gained enormous popularity and attention; however, these tools are often unreliable
\citep{imai2022github,
nguyen2022empirical, wermelinger2023using}, e.g., failing to understand users' intentions~\citep{barke2023grounded} and generating bugs that may be particularly risky alongside beginner programmers~\citep{dakhel2023github}. Programming involves much more than just accurate in-line code suggestions -- which, at the time of writing, GitHub Copilot specializes in. Humans plan abstract, structural decisions and collaboratively learn, 
and need partners who can answer our questions -- like \textit{why} code behaves as it does, or fails to work.
A good collaborative programming partner seeks to understand not only the programming language, but also their fellow \emph{programmer}, inferring and reasoning about our overarching intentions, and adapting to both what we do and do not know.

\textit{Thought Partners for Embodied Assistance.} Ensuring embodied agents can form accurate and physically-realizable plans is foundational for effective assistance we can trust -- from guessing what a friend wants when we help them cook \citep{fisac2020pragmatic}, to working with someone with different physical abilities \citep{ranz2017capability}, or carrying out a high-stakes search-and-rescue mission \citep{casper2003human}. While much current research on embodied AI and assistive robots focuses on learning specific skills or following simple instructions \citep{shridhar2020alfred,ahn2022can,raad2024scaling}, evaluations suggest that even state-of-the-art language models fine-tuned on extensive human feedback continue to struggle with tasks that require reliable, effective planning towards novel goals \citep{valmeekam2022large,momennejad2024evaluating}. Instead, ideal assistive partners understand our actions, words, and instructions as expressions of goals, beliefs, and intentions \citep{goodman2016pragmatic,sumers2023reconciling,jeon2020reward} that are \emph{grounded} in physical possibilities \citep{kollar2013generalized}, while also understanding that these can be \emph{shared} across multiple minds \citep{bratman2013shared,stacy2021modeling,wu2021too}. In addition, effective partners account for each others' limitations in perception, planning, and world modeling, correcting for possible mistakes \citep{reddy2018you,alanqary2021modeling}, and acting so as to make their intentions more legible  \citep{dragan2013legibility,miura2021unifying}.

\textit{Thought Partners for Storytelling.} Another domain in which we may want thought partners is storytelling -- for writers, filmmakers, and even scientists. Storytelling is a complex, iterative cognitive process~\citep{flower1981cognitive,hayes2012modeling} with substantial opportunities for thought partners to collaboratively ideate and create with humans from helping brainstorm new ideas, generate storylines, and improve their writing style and tone~\citep{lee2022coauthor,lee2024design,ippolito2022creative,gero2022sparks,gero2023social,dell2023navigating}. For this process to be productive, a thought partner needs to understand more than just our authorial intentions and dispositions -- they also need to understand the \emph{audience} we are speaking to (that is, to \emph{understand the social world}), including audience expectations and likely interpretations of the stories we are crafting for them.

\textit{Thought Partners for Medicine.} Doctors need to sensemake, plan, deliberate, and continually learn in the face of new medical evidence. A primary care doctor is not unlike Sherlock Holmes -- collating and integrating disparate bits of evidence with their prior beliefs to make decisions under uncertainty. Yet, doctors rarely have enough time to engage deeply with each patient~\citep{porter2023revisiting}, driving high rates of burnout with knock-on effects on patient care quality~\citep{dewa2017relationship}. Can we develop safe, reliable thought partners that can free doctors up to spend more time and communicate better with their patients? Already, foundation models are becoming proficient in medical assessments~\citep{chowdhery2022palm, singhal2023large}, seemingly capable of easing the heavy burden on doctors by assisting and partnering~\citep{topol2019high, tu2024conversational}, and even providing preferable responses to patients~\citep{chatbotMed}. Yet, it is not clear that these systems \textit{understand us} (and our cognitive limitations), \textit{understand the world} (underlying biology), and enable us to \textit{understand them} (which in this context, may be important for transparency and reliability~\citep{vellido2020importance, rajpurkar2022ai, ghassemi2020review, daneshjou2021lack}).

\subsection{Desiderata}
\label{desiderata}

What then do we want from thought partners? There are many criteria for tools for thought that are of course relevant: efficiency, accuracy, robustness, fairness, cost, scalability, etc. But the domains above illuminate that what is distinctive about a thought \emph{partner} is its relationship to the \emph{user}~\citep{cabitza2019proof}. Looking to ideas the behavioral sciences motivates three desiderata to guide the design of human-centered thought partners:

\begin{enumerate}
    \item \textbf{You understand me}: We would like our thought partners to understand our goals, plans, (possibly false) beliefs, and resource limitations, taking into account what they have observed of us in the past and present in order to best collaborate with us in the future~\citep{puig2020watch, chandra2024inferring}. For example, a thought partner should adaptively change strategies when working with an expert, layperson, or child, meeting us where we are.
    \item \textbf{I understand you}: We would like our thought partners to act in a way that is legible to us \citep{dragan2013legibility, fisac2020generating}, and communicate with us in the way we intuitively understand \citep{grice1975logic, doshi2017towards, miller2019explanation}.
    \item \textbf{We understand the world}: We would like our thought partners to be tethered to reality \citep{smith2019promise}. This means being accurate and knowledgeable, but also working with a shared representation of the world, domain, or task \citep{sucholutsky2023alignment, sucholutsky2023getting, battaglia2013simulation}. Further, our use of `\textit{we}' emphasizes that thought partnerships are fundamentally about \textit{synergy}, moving beyond the sum of its parts.

\end{enumerate}

\begin{table}[ht!]
\small
\caption{\textbf{Modes of collaborative thought.} Settings in which human-human and human-AI thought partners can engage.}
\label{tab:modes_of_thought}
% \begin{tabular}{@{}llll@{}}
\begin{tabular}{p{4.5cm}p{4.75cm}p{5.25cm}}
\toprule
\textbf{Mode}                           & \textbf{Ongoing Challenges}               & \textbf{Sampling of Existing Systems}       \\
\midrule 

\begin{tabular}[c]{@{}p{4.25cm}@{}}Collaborative planning  \begin{itemize}[leftmargin=*,nosep]
   \item Joint decision-making
   \item Decentralized cooperation
   \item Goal and task assistance
\end{itemize}\end{tabular}         & \begin{tabular}[c]{@{}p{4.75cm}@{}}Reliable goal inference\\ Value and intent alignment\\ Scalable multi-agent planning\end{tabular}  &
\begin{tabular}[c]{@{}p{5.25cm}@{}}
Collaborative robots \citep{dragan2013legibility,roncone2017transparent} \\
Video game sidekicks \citep{carroll2019utility,macindoe2012pomcop} \\
Language-based assistants \citep{zhi2024pragmatic,lin2022inferring}
\end{tabular}
\\
\midrule
\begin{tabular}[c]{@{}p{4.25cm}@{}}Collaborative learning \begin{itemize}[leftmargin=*,nosep] \item Pair \& team problem-solving \item Identification of knowledge gaps \item New problem construction \end{itemize}\end{tabular} & \begin{tabular}[c]{@{}p{4.75cm}@{}} Strong \& robust problem-solving abilities  \\ Personalized curriculum pacing \\ Problem construction of targeted difficulty 
% Assessment of educational content reliability
\end{tabular}  & \begin{tabular}[c]{@{}p{5.25cm}@{}}

Programming learning aids~\citep{keuning2018systematic,  sarsa2022automatic, chandra2024watchat, head2015tutorons} 
\newline 
Mathematics tutors ~\citep{rafferty2020assessing, collins2024evaluating, poesia2023peano}       \end{tabular}                                                                                                 \\
\midrule 
\begin{tabular}[c]{@{}p{4.25cm}@{}}
Collaborative deliberation  \begin{itemize}[leftmargin=*,nosep]    \item Debate \& argumentation \item Critical review \& discussion \item Consensus formation\end{itemize}       \end{tabular}                   & \begin{tabular}[c]{@{}p{4.75cm}@{}}Opinion diversity\\ 
Verifiable reasoning \\ Formation of common ground\end{tabular}     & \begin{tabular}[c]{@{}p{5.25cm}@{}}Machine-assisted debating \citep{slonim2021autonomous, jarrett2023language, du2023improving}\\
Consensus writing \& opinion mapping \citep{bakker2022fine, small2021polis} \\
\end{tabular}                                                                    \\
\midrule  \begin{tabular}[c]{@{}p{4.25cm}@{}}
Collaborative sensemaking  \begin{itemize}[leftmargin=*,nosep] \item Explanation \item Visualization \item Data Analytics \end{itemize} \end{tabular} & \begin{tabular}[c]{@{}p{4.75cm}@{}}Exponential increases in data produced \\ Accessible communication \\ Fidelity of insights to the world\end{tabular} & \begin{tabular}[c]{@{}p{5.25cm}@{}} Probabilistic data modeling~\citep{lew2021pclean, saad2019bayesian, huot2024gensql, li2024automated, steinruecken2019automatic} \\
Machine-assisted theory discovery \citep{davies2021advancing, cranmer2020discovering, romera2024mathematical}
\end{tabular}  
\\
\midrule 

\begin{tabular}[c]{@{}p{4.25cm}@{}}Collaborative creation \& ideation \begin{itemize}[leftmargin=*,nosep] \item Co-design \item Idea critiquing \item Brainstorming\end{itemize}\end{tabular}                                               & \begin{tabular}[c]{@{}p{4.75cm}@{}} Generation diversity\\Style consistency\\ Modular customizability\end{tabular}                  & \begin{tabular}[c]{@{}p{5.25cm}@{}}Machine-assisted writing\\\citep{lee2022coauthor, ippolito2022creative, ashkinaze2024ai}\\ 
Prompted image creation~\citep{suri2024the, vartiainen2023using, gafni2022make} \\
Collaborative sketching~\citep{fan2019collabdraw, ge2020creative, dvorovzvnak2020monster}
\end{tabular}                                                                                           \\

\bottomrule
\end{tabular}
\end{table}

\begin{table}[ht!]
\small

\caption{\textbf{Bayesian Thought Partner Toolkit.} A range of \textit{computational cognitive motifs} for reverse engineering the mind in engineering terms, drawn from computational cognitive science, can be used to build human-centric thought partners that meet our desiderata. 
}
\label{tab:toolkit}
% \begin{tabular}{@{}llll@{}}
% \begin{tabular}{p{4.5cm}p{4.25cm}p{4.25cm}}
\begin{tabular}{p{2.5cm}p{7.5cm}c}
\toprule
\textbf{Motif}                                                                                                     & \textbf{Description}               & \textbf{Sample References}%Sampling of Key References}                               

\\ \midrule 

\begin{tabular}[c]{@{}p{2.5cm}@{}}Probabilistic Mental Models and Inference \end{tabular}                            & \begin{tabular}[c]{@{}p{7.25cm}@{}}Humans update beliefs and draw inferences consistent with probabilistic generative models representing the world.\end{tabular}    & \begin{tabular}[c]{@{}p{4.25cm}@{}}\centering ~\citep{chater2008probabilistic, griffiths2008bayesian, tenenbaum2011grow}
\end{tabular} 

\\ \midrule

\begin{tabular}[c]{@{}p{2.5cm}@{}}Structured Knowledge Representations \end{tabular}                            & \begin{tabular}[c]{@{}p{7.25cm}@{}}Humans have abstract, highly structured conceptual representations that include causality, agents, and physical representations.%\textit{Thought partners can build on representations that explicitly acquire and maintain structured and causal knowledge.}
\end{tabular}    & \begin{tabular}[c]{@{}p{4.25cm}@{}}\centering ~\citep{spelke2000core,piantadosi2021computational,quilty2023best}
\end{tabular} 

\\ \midrule 

\begin{tabular}[c]{@{}p{2.5cm}@{}}Hierarchical Models \end{tabular}                            & \begin{tabular}[c]{@{}p{7.25cm}@{}}Humans construct and update \textit{hierarchical} representations that separate concrete knowledge and belief from abstract ones.%\textit{Thought partners can build on explicitly hierarchical representations to learn human-like abstract hypotheses from concrete observations.}
\end{tabular}    & \begin{tabular}[c]{@{}p{4.25cm}@{}} \centering ~\citep{griffiths2003hierarchical,kemp2008discovery,lake2015human}
\end{tabular} 

\\ \midrule 

\begin{tabular}[c]{@{}p{2.5cm}@{}}Theory Learning as Program Synthesis \end{tabular}                            & \begin{tabular}[c]{@{}p{7.25cm}@{}} Humans minds can be viewed as growing and editing theories of the world, expressed as programs, to ``improve'' their codebase (world models).
\end{tabular}    & \begin{tabular}[c]{@{}p{4.25cm}@{}} \centering ~\citep{rule2020child, ellis2021dreamcoder,ullman2020bayesian}
\end{tabular} 

\\ \midrule 

\begin{tabular}[c]{@{}p{2.5cm}@{}}Resource-Rationality \end{tabular}                            & \begin{tabular}[c]{@{}p{7.25cm}@{}}
Humans make rational choices about how to allocate finite computational resources, including time and memory.\end{tabular}    & \begin{tabular}[c]{@{}p{4.25cm}@{}} \centering ~\citep{lieder2019cognitive, lieder2020resource, icard2023resource}
\end{tabular}

\\ \midrule 

\begin{tabular}[c]{@{}p{2.5cm}@{}}Goal-Directed Planning and Search \end{tabular}                            & \begin{tabular}[c]{@{}p{7.25cm}@{}}
Humans are intentional actors, who plan to achieve goals by reasoning about the (uncertain) effects of their (possible) actions in the environment. \end{tabular}    & \begin{tabular}[c]{@{}p{4.25cm}@{}} \centering ~\citep{newell1972human, mattar2022planning, ho2022people}
\end{tabular}

\\ \midrule 

\begin{tabular}[c]{@{}p{2.5cm}@{}} Bayesian Theory of Mind (BToM) \end{tabular}                            & \begin{tabular}[c]{@{}p{7.25cm}@{}} Humans represent \textit{other} agents as intentional, intelligent actors; and probabistically infer their mental states from observations of actions. \end{tabular}    & \begin{tabular}[c]{@{}p{4.25cm}@{}} \centering ~\citep{jara2016naive, baker2011bayesian, ho2022planning}
\end{tabular} 

\\ \midrule 

\begin{tabular}[c]{@{}p{2.5cm}@{}}Rational Speech Acts (RSA) \end{tabular}                            & \begin{tabular}[c]{@{}p{7.25cm}@{}}Humans reason about language as an intentional, communicative action to infer a speakers' underlying goals.\end{tabular}    & \begin{tabular}[c]{@{}p{4.25cm}@{}} \centering ~\citep{frank2012predicting, goodman2016pragmatic, degen2023rational}
\end{tabular}  

\\ \midrule 

\begin{tabular}[c]{@{}p{2.5cm}@{}}Learning to Learn\end{tabular}                            & \begin{tabular}[c]{@{}p{7.25cm}@{}} Humans \textit{meta-learn} (improve our overarching ability to learn) jointly with learning new concrete concepts and skills. 
\end{tabular}    & \begin{tabular}[c]{@{}p{4.25cm}@{}}\centering ~\citep{binz2023meta, grant2018recasting, lake2023human, lake_ullman_tenenbaum_gershman_2017}
\end{tabular}

% \end{tabular}

% \end{table}
\\

\bottomrule
\end{tabular}
\end{table}

\section{Engineering Human-Centered Thought Partners }
\label{design-thought-partners}

Our core proposal is that our three desiderata can be engineered explicitly, building on theoretical motifs from computational cognitive science and cognitively-informed AI (summarized in Table \ref{tab:toolkit}), rather than left as emergent and potentially brittle properties arising implicitly in systems trained for other ends~\citep{mccoy2023embers}. Here, we articulate a framework for engineering thought partners designed to robustly and explicitly function as cooperative, collaborative actors. Humans are far from homogeneous, perfectly rational oracles, nor are we so unpredictable that it is hopeless to model human behavior. We argue that models that explain human cognition and choice as approximately optimal solutions given goals and constraints provide an ideal starting point for designing thought partners, and that a Bayesian formalism provides a probabilistically-sound common conceptual language that facilitates cross-talk between different disciplines~\citep{griffiths2023bayes,ho2022cognitive, yang2023inner}. 

\subsection{Implementing Our Desiderata}

What does it take to engineer real systems that meet our desiderata? First, we propose that a thought partner that understands us should explicitly \textit{model its human collaborator} as such -- as a cooperative agent with structured internal beliefs, knowledge, and goals -- and fundamental resource limitations. Second, engineering a thought partner that we can understand benefits from looking at how \textit{humans model other humans}; just as a good human collaborator seeks to learn and adapt to the relative strengths, imperfections, and computational bounds of their partner, we can build machine thought partners that also reason about the computational demands they are placing on another agent such that we can appropriately predict their behavior~\citep{gweon2023socially, steyvers2023threePublished}. Finally, to build thought partners that understand the world -- and learn and think synergistically alongside us -- we argue that it is valuable to build on structured computational toolkits for \textit{grounding shared goals and communication into the environment and domain} in which collaboration takes place.  

\subsection{Computational Cognitive Science Motifs}

We now (non-exhaustively) spotlight several key insights about modeling humans, modeling humans modeling humans, and modeling humans modeling the world from computational cognitive science -- ``motifs'' for reverse engineering the mind (Table \ref{tab:toolkit}) -- that we believe can inform engineering of human-centered thought partners. While we acknowledge that there are communities within cognitive science that may disagree with some of these theories, we emphasize that the computational underpinnings of the motifs hold tremendous engineering potential for building thought partners in practice. 

\textit{Probabilistic Models of Cognition.} Decades of work in computational cognitive science have demonstrated the power of modeling aspects of human cognition as Bayesian inference through structured probabilistic generative world models \citep{griffiths2008bayesian, chater2006probabilistic, oaksford2007bayesian, tenenbaum2011grow, lake2015human}. Such approaches have found empirical success in capturing a diversity of facets of human cognition from early word learning \citep{xu2007word}, to visual perception~\citep{kersten2004object, yildirim2020efficient}, physical reasoning~\citep{battaglia2013simulation, allen2020rapid, zhang2023grounded}, concept learning \citep{tenenbaum1998bayesian, goodman2008rational, piantadosi2016logical}, language processing and acquisition~\citep{chater2006probabilistic, griffiths2004integrating, goodman2015probabilistic, yang2022one}, causal inference in children~\citep{schulz2007can, gopnik2004theory} and adults~\citep{kirfel2022inference, lagnado2013causal}, memory reconstruction~\citep{hemmer2009bayesian}, and theory formation~\citep{ullman2020, griffiths2009theory}, among many others. Probabilistic models of cognition, particularly those built using a Bayesian approach, have offered principled formalisms in capturing rapid belief updating~\citep{vul2014one} and how we may integrate our commonsense world knowledge with new evidence to inform the actions and decisions we take in the world~\citep{ho2022planning}. Probabilistic inference over structured representations, particularly drawing on Bayesian modeling and tools like meta-level Markov Decision Processes~\citep{hay2014selecting}, has provided a computational account of how humans plan so flexibly, with the capability of forming rich hierarchical goals and subgoals, across varied timescales \citep{ho2022planning, ho2022cognitive, tomov2020, baker2014modeling, callaway2022rational}. %The engineeing 

\textit{Theory of Mind and Communication.} In our quest to build systems for collaborative cognition, we are guided by the success of Bayesian accounts of how we reason about \emph{others}' mental states, and how we communicate about them. In particular, Bayesian treatments of theory of mind (ToM) have offered strong accounts for how we may rapidly reason about each others' beliefs, desires, goals, and intentions \citep{baker2009action, baker2017rational, zhi2020online, ying2023neuro, jara2016naive}. We may build mental models~\citep{johnson1983mental, byrne2002mental} of our thought partners, which can in turn be used to support communication and collaboration, informing the way we teach~\citep{shafto2014rational, sumers2021learning, liquin2023teaching}, infer whether to rely on a partner for help~\citep{kumar2023differentiating}, and support rapid, flexible adaptation to new conversation partners~\citep{hawkins2023partners, hawkins2023flexible}. We call particular attention to the Rational Speech Act (RSA) framework \citep{frank2012predicting, goodman2016pragmatic}, which models communicative partners as recursively reasoning about each others' minds to inform what to say (from the perspective of the speaker) and how to interpret a received utterance (as the listener). Bayesian models provide a useful framework for formalizing such rich cross-partner inferences, allowing both social cognition and communication to be modeled with the same computational toolbox \citep{goodman2013knowledge,ho2021communication}. %\K{clarify the connection between theory of mind and communication. both can be modeled as a kind of multiplayer game with the same computational toolbox.} %and inform thought-partner design. % (see Sections \ref{bayesian-hmt-proposal} and \ref{resource-rational}).

\textit{Resource-Rationality and Tractable Theory-Building.} Human brains also have limited resources such as time, memory, and attention that shape what we think about, how long we spend thinking, and even how we communicate our thoughts to others~\cite{griffiths2020understanding}. Thus, we sometimes make systematically biased inferences~\citep{tversky1973availability, tversky1974judgment}. Such ``erroneous'' judgments can be captured by modeling humans as making rational use of our finite resources; e.g., via approximate inference ~\citep{vul2014one, zhu2023autocorrelated} or bounded planning \citep{alanqary2021modeling}. Crucially, human cognition is tractable~\citep{van2008tractable}. Indeed, we can navigate large, potentially unbounded, hypothesis spaces to build theories of the world: a process that seems to demand some kind of heuristics and approximations, which may be resource-rational ~\citep{callaway2022rational, icard2015resource, griffiths2020understanding, icard2023resource, lieder2020resource, anderson1990adaptive, zhang2023ai}. One approach to modeling minds advocates thinking about humans, as ``world model builders'' (or ``hackers'') -- conducting experiments and updating our beliefs about compressed representations of the world, where these representations may be expressed as programs~\citep{rule2020child, ullman2020}.
Such representations -- bolstered by tools like program synthesis -- help explore suboptimal behavior~\citep{cheyette2023people}.

\subsection{Scaling Thought Partners via Probabilistic Programming} 

If Bayesian thought partners are to reason over models of their human thought partner and the world, these models need to continually evolve as new facts come to light and as the human thought partner themselves grows in their expertise, beliefs, and needs. Probabilistic programming ~\citep{goodman2008church} provides one powerful methodology for building, scaling, and performing inference in these kinds of rich models. For example, probabilistic programs can be learned from data ~\citep{saad2022scalable, saad2019bayesian}, and synthesized via LLMs that encode rich priors ~\citep{wong2023word, li2024automated, lew2020leveraging}. Probabilistic programs also enable fast approximate inference in world models that cohere with human common-sense knowledge and domain expertise \citep{lew2021pclean, gothoskar2023bayes3d}, where the learned models are themselves amenable to modular inspection and editing by humans. Modern probabilistic programming languages ~\citep{cusumano2019gen, bingham2019pyro, mansinghka2018probabilistic} offer not just generic inference but \textit{programmable} inference, that is, they automate the math for hybrids of optimization \citep{adevLew2023, becker2024probabilistic}, dynamic programming \citep{saad2021sppl}, and Monte Carlo inference \citep{lew2023probabilistic}. While such frameworks are certainly not the only methods to handle uncertainty and build effective and robust thought partners, we believe they are one promising and cognitively-grounded approach to instantiating thought partners today, as we discuss in our case studies.

\subsection{Infrastructure around Thought Partners}

The design of systems that learn and think with people necessitates not only careful construction of the thought partner (i.e., the machine itself), but also the \textit{infrastructure} within which human and computational thought partners collaborate~\citep{steyvers2023threePublished}. Questions like ``when and where should a human be able to engage a computational thought partner to ensure effective and appropriate use?'' or ``for a given problem, is the human or computational thought partner better suited to start first, in light of their respective strengths and weakness, costs of the task at hand, and particular mode of thought?'' inform the design of the workflow that \textit{surrounds} thought partnership. This sociotechnical ecosystem may be dictated by external regulations, organizational practices, or other principles \citep{guggenberger2020ecosystem,lee2024design, goodman2017european,wachter2019right,fui2023generative}, and crucially informed by studies of human behavior. For example, Article 14 of the EU AI Act requires users of high-risk AI systems ``to correctly interpret the high-risk AI system’s output'' and ``to remain aware of the possible tendency of automatically relying or over-relying on the output.'' Satisfying such requirements begets not only careful design of thought partners (e.g., that we can understand), but demands careful design of the system of affordances~\citep{norman1988design, chemero2018outline} and infrastructure around thought partnerships (for instance, communicating back to humans information about their reliance strategies).  Disentangling thought partners from the infrastructure around them provides a \textit{modular} scaffold for addressing unintentional thought partnership behavior, e.g, overreliance~\citep{zerilli2022transparency} and ``illusions of understanding''~\citep{messeri2024artificial}. Bayesian modeling has already found success in inferring humans' reliance strategies~\citep{tejeda2022ai} and regions of the task space where a human versus machine can complement one another~\citep{steyvers2022bayesian}.

\section{Case Studies in Engineering Thought Partners}

\begin{figure}[t!]
    \centering
    \includegraphics[width=1.0\linewidth]{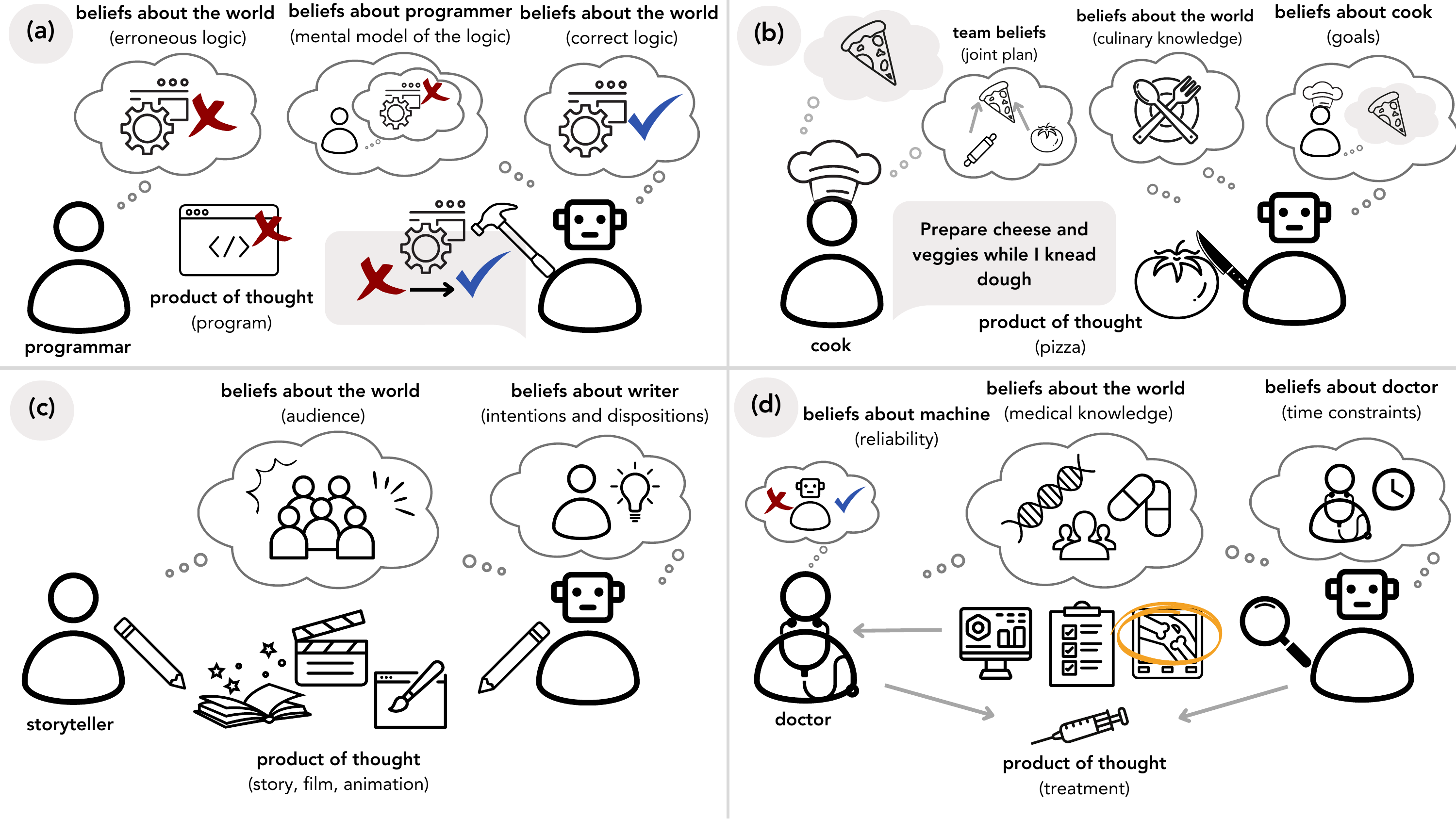}
    \caption{\textbf{Case Study Depictions.} (a) WatChat infers the user's buggy mental model of the programming environment and interactively helps ``patch'' bug(s) in their understanding; (b) CLIPS reasons explicitly about agents' goals, integrating (culinary) world knowledge and the human's utterances to infer appropriate actions. Both agents reason about the \textit{joint team plan} (tomato and dough are needed to make pizza); (c) Thought partners based on inverse inverse storytelling explicitly reason over models of the audience; (d) Future thought partners for medicine can jointly reason with a human doctor across modalities, a shared understanding of biology and patient needs, and a model of others' limitations.}
    \label{fig:case-studies}
\end{figure}

We now return to the example domains previously introduced and discuss specific case studies (depicted in Figure~\ref{fig:case-studies}). Our goal is to demonstrate the potential benefits of endowing thought partners with structured probabilistic models of the human and/or world, and provide a flavor of the kinds of infrastructure questions that may surround them to ensure that the thought partners we build work \textit{with} people. 

\subsection{Thought Partners for Programming}

We highlighted some visions for effective programming partnerships, such as a partner that can address ``why'' questions. 
One recent idea, from \citet{chandra2024watchat}, is to apply the Bayesian toolkit to \emph{explain surprising behavior} of computer programs in a human-like way. \citeauthor{chandra2024watchat}, 
apply Bayesian models of mental state inference and rational communication \citep{chandra2024cooperative} to design a system called ``WatChat'' that answers questions like ``why did program $p$ output result $r$?'' in a principled, human-like way. 
WatChat infers what \emph{erroneous mental model} might cause the programmer to have expected something different (partner understands user) and generates an explanation that ``debugs'' that mental model (user understands partner). WatChat represents possible mental models themselves as ``programs'' whose ``bugs'' correspond to possible misconceptions; mental models can thus be inferred by Bayesian program synthesis (see Table~\ref{tab:toolkit}). Such a framework can also be inverted to help design new questions for teachers or self-driven learners to identify misconceptions.

\subsection{Thought Partners for Embodied Assistance} 

Recall the challenge of collaboratively planning uncertain tasks, from a search-and-rescue mission to everyday cooking, wherein we typically want to infer shared goals and communicative intent from our partners. This cooperative logic can be modeled in a Bayesian architecture called Cooperative Language-Guided Inverse Plan Search (CLIPS) \citep{zhi2024pragmatic}. By modeling humans as cooperative planners who use language to communicate \emph{joint plans} to achieve their goals \citep{wu2021too}, CLIPS is able to infer those plans and goals from both the actions and instructions of human collaborators. This allows CLIPS to \emph{pragmatically} follow human instructions, using context to disambiguate the multiple meanings that a request might have, while \emph{pro-actively} assisting with the goals that underlie the instruction. For example, CLIPS can understand the likely intentions behind an instruction like ``Can you prepare the vegetables while I knead the dough?'', inferring the shared goal of making pizza. These capabilities are made possible by using probabilistic programming infrastructure \citep{cusumano2019gen} to unite algorithms for Bayesian inverse planning \citep{baker2009action,zhi2020online} and human-AI alignment  \citep{hadfield2016cooperative,fisac2020pragmatic,jeon2020reward} with LLMs. In particular, by using LLMs to evaluate the probability of a natural language instruction given a possible intention, CLIPS can infer intentions from natural language in a coherent Bayesian manner -- demonstrating the power of combining tools from the Bayesian thought partner toolkit.

\subsection{Thought Partners for Storytelling}
Storytelling is about crafting \textit{experience}. Can we also apply the toolkit to help storytellers design experiences from first principles? Recent work has shown that a system grounded in Bayesian ToM can predict and even design interventions on the audience's experience of a story~\citep{chandra2023acting, chen2024intervening}. \citeauthor{chandra2024storytelling} conceive of storytelling as ``inverse inverse planning'': that is, starting with human social cognition, modeled as Bayesian inverse planning \citep{baker2009action}, and then optimizing narrative events to shape the model's inferences over time. They show how a variety of storytelling techniques -- from plot twists to stage mime -- can be expressed in the language of inverse inverse planning to create animations that have a desired cognitive effect on viewers. Herein, we also highlight the breadth of thought partners for media beyond language, though the framework does nicely suggest a variety of natural extensions, such as integration into tools for creative writing \citep{lee2022coauthor,lee2024design,ippolito2022creative,gero2022sparks,gero2023social,dell2023navigating}.

\subsection{Thought Partners for Medicine}

Finally, we envision medical thought partners both understand us -- reasoning about the doctor, patient, and care team as agents with goals, beliefs, and worries -- and \textit{complement} our capabilities, integrating swaths of evidence that exceed our cognitive capacities to inform diagnosis and treatment. While no system yet meets our desiderata for these criteria, we believe a range of motifs and tools from the Bayesian thought partner toolkit here can support the development of such systems for collaborative sensemaking and deliberation. We imagine Bayesian thought partners that can update their medical world knowledge in light of new insights in biology, e.g., editing a code snippet of the underlying probabilistic world model~\citep{wong2023word} or growing the representation in a non-parametric hierarchical Bayesian model~\citep{griffiths2003hierarchical}. Such a model can then, similar to WatChat, synthesize new questions to ensure the human doctor's own medical world model is sound. 
Early work demonstrates that we can employ elements of our toolkit, specifically probabilistic programming, to learn rich generative models for oncology and support efficient user queries~\citep{loula2024learning}. Yet, effective medical thought partners beckon a broader view of the ecosystem in which they are deployed~\citep{cabitza2019proof, cabitza2017unintended}. If a doctor is over-relying on the output of the thought partner, or overburdened amidst a surge in patient queries, infrastructure around the human and thought partner can modulate when a patient query is either routed to a human or the AI thought partner, or deemed necessary of collaborative planning~\citep{mozannar2020consistent}. Systems for routing based on probabilistic modeling are already proving successful in simulation~\citep{dvijotham2023enhancing}. 
 
\section{Looking Ahead}

There is much exciting work to be done to characterize when and how to build thought partners across modes of collaborative thought, which can advance the dissemination and creation of new knowledge alongside humans. We next lay out several key challenges for researchers and designers intent on pursuing a human-centered program of building machines that learn and think \textit{with} people. 

\subsection{Non-Dyad Settings}

While there is substantial work to be done characterizing the space of possibilities for a single human and single AI thought partner (``dyadic''), we envision a future where \textit{many} humans and \textit{many} machines engage (``non-dyadic''), across roles and specialties in increasingly complex social systems~\citep{tsvetkova2024human}, engage in the realm of thought~\citep{schneiders2022non, hornecker2022beyond, yadav2024beyond}. Already, researchers are exploring non-dyadic versions of many of the modes of thought and case studies laid out above, including collaborative learning with groups of humans accompanied by an AI thought partner~\citep{ sucholutsky2024representationalTeaching} and medical robot collision avoidance systems that need to account for multiple humans \citep{li2023three}. As in the dyad setting, extensions to non-dyadic settings can be bolstered by a deepening understanding of human behavior in groups -- expanding the Bayesian thought partner toolkit -- as is already underway in the study of convention formation~\citep{hawkins2023partners, boyce2024interaction}. Looking ahead, citizen science is a promising example of the opportunities of creating large networks of humans and thought partners: Zooniverse, a large-scale galaxy classification crowdsourcing project, serves as a case study for exploring smart task allocation, blending human and machine classifications, and infrastructure changes that impact human participation and performance with outcomes including both iterative scientific progress and serendipitous scientific discovery \citep{trouille2019citizen}.

\subsection{Evaluation}

The assessment of thought partners demands a multi-faceted, cross-disciplinary suite of approaches. At minimum, the evaluation of AI thought partners must include some element of \textit{interactivity}~\citep{hornbaek2017interaction}. Recent works have highlighted deficits in static evaluation of foundation models~\citep{lee2023evaluating, collins2024evaluating}, demonstrating the need for considering the interaction \emph{process} in addition to
the final output, the \emph{first-person} perspective in addition to the third-party perspective, and notions of \emph{preference} beyond quality. In addition to interactive user studies, we posit that to study different kinds of thought partners across modes of collaborative thought would benefit from a controlled, yet rich, playspace; \textit{games} provide one such domain. Games offer a good formalism for the study of repeated interactions between multiple agents and grounds to explore rich patterns of thought, in social collaborative settings~\citep{allen2024using, park2023generative, brown2019superhuman, meta2022human}. 

\subsection{Risks and Important Considerations}

Computational thought partners are by no means a guaranteed nor universal good and come with certain risks. We call out three such spheres of risk: (i) reliance, critical thinking, and access, (ii) anthromorphization, and (iii) misalignment.

First, AI thought partners could induce over-reliance and impair the development of critical thinking skills~\citep{logg2019algorithm, green2019principles, inuwa2023algorithmic, messeri2024artificial}, potentially acting as ``steroids'' for the mind ~\citep{hofman2023steroids}. We are concerned about these risks; our emphasis on the \textit{infrastructure} around thought partner use is explicitly intended to help practitioners take steps to address these challenges, motivating further design of infrastructure modifications like cognitive forcing functions~\citep{buschek2021impact,buccinca2021trust}. Conversely, it is possible that some people may \textit{under-rely} on a thought partner, particularly if there is inadequate AI literacy training for how to best make use of new thought partners~\citep{dietvorst2015algorithm,dietvorst2018overcoming,zerilli2021how}. Already, research has found that the kinds of queries people make of AI systems can be informed by the amount of prior experience they have interacting with chatbots~\citep{collins2024evaluating} meaning students, researchers, and other practitioners in lower-income communities may be unable to maximize the value of thought partnering. It is important to ensure that the benefits of thought partners are not confined to an exclusive set of people. 

Second, on the topic of anthromoprhization, we highlight an important distinction between \textit{human-centric} and \textit{human-like} thought partners~\citep{mumford2010technics}. Our desiderata ``I understand you'' advocates for thought partners whose behavior we understand; while this could draw on how we understand other humans, however, we should be careful about \textit{interpreting} such machine thought partners \textit{as} we do humans. As \citet{weizenbaum1966eliza} illuminated with the ELIZA system, there are risks to developing computer systems that present themselves as human-like in ways that they are not: for example, by leading users to attribute undue intention to systems' responses or (in the long run) leading society to devalue human intelligence \citep{weizenbaum1976computer}. Human-like thought partners should maintain categorical delineation between humans and machines to prevent overreliance~\citep{logg2019algorithm,weidinger2022taxonomy} and promote human dignity without encroaching on any partner's self-worth~\citep{shneiderman2022human}. 
The term used to refer to a thought partner can affect the assumptions made about their capabilities  (e.g., \emph{teammate} implies the machine and human are on equal footing) or can detract from a partner's human-like nature (e.g., \textit{tool} would be less anthropomorphic).

Lastly, we note that insufficiently accurate, robust, or cognitively-grounded models can yield \emph{misalignment} with humans, leading intended AI thought partners to act towards the wrong goals \citep{zhuang2020consequences}, provide wrong or misleading information \citep{kalai2023calibrated}, or violate safety constraints \citep{amodei2016concrete}. A Bayesian approach to thought partnership can address some of these issues, enabling uncertainty-aware decision-making that avoids overconfidence \citep{hadfield2016cooperative, russell2019human, russell2021artificial}. Yet, while inferring human thoughts and behavior can be used to design better collaborators, models of humans are inherently dual-use and can also be used to mislead, surveil, or manipulate \citep{carroll2023characterizing}. It is crucial to consider whether thought partners are aligned with society at large, or merely superficially aligned with users while serving more powerful interests \citep{lazar2023ai}.

\section{Conclusion}

If we are to build helpful and reliable human-AI thought partnerships, we advocate for design that explicitly recognizes and engages with the richness and diversity of human thought in an often unpredictable world. We have argued, supported by several case studies, that those engineering thought partners and the infrastructure around their use can benefit from drawing on motifs from computational cognitive science and cognitive-AI. The future of collaborative cognition is bright, but not without risk; continual collaboration and knowledge sharing amongst behavioral scientists, AI practitioners, domain experts, and related disciplines is crucial as we strive to build machines that truly learn and think \textit{with} people.

\clearpage
%TC:ignore

 \noindent\fbox{%
    \parbox{\textwidth}{%
    
        \textit{Glossary of main terms}

\begin{itemize}
    \item Collaborative cognition: the process by which two or more agents work together in some aspect(s) of thinking (e.g., planning together, learning together, creating together). 
    \item Thought partner: another entity (human or AI) that works with an agent to push forward some aspect(s) of thinking. 
    \item Artificial Intelligence (AI): computational systems that are able to process inputs and engage in some aspect of learning, planning, reasoning, and/or decision-making. Used interchangeably with machines. 
    \item Large language model (LLM): a particular kind of AI system which learns a distribution over text, often trained on large amounts of web-scale text data. LLMs are a class of large-scale foundation models. 
    \item Agent: an entity that can process inputs, make decisions, and take actions in some environment.
    \item Dyad: a system with two agents (e.g., human-human, human-AI, AI-AI). 
    \item Resource-rationality: the idea that human behavior and cognition can be viewed as \textit{rational} under bounded constraints (e.g., under limited working memory).
    \item Probabilistic generative model: a model of how the data one observes about the world is generated by some probabilistic process, from which one can sample new observations and make queries about existing observations.
    \item Probabilistic programming language (PPL): a language for expressing probabilistic generative models as computer programs that interleave deterministic code (e.g. arithmetic, logic, or artificial neural networks) with random choices. PPLs allow users to specify probabilistic models and inference algorithms in a modular and compositional manner.
    \item Bayesian inference: a method for updating one's \textit{beliefs} over various aspects of the world, grounded in probability theory; in Bayesian inference, an agent updates their beliefs by assigning higher credence to hypotheses that better explain the evidence, weighted against the backdrop of their prior beliefs.
    \item Affordance: design features of a system that inform use.
\end{itemize}
}
}
%TC:endignore

%TC:ignore

\section*{Acknowledgments}

We thank Richard Turner, Laura Schulz, Tyler Brooke-Wilson, Valerie Chen, Alena Rote, Lance Ying, Tony Chen, Matt Ashman, Mike Walmsley, Albert Jiang, Mateja Jamnik, Dj Dvijotham, Jonathan Ragan-Kelley, Will Crichton, Alex Lew, Tim O'Donnell, Joao Loula, Marty Tenenbaum, Mary McNaughton-Collins, and Jim Collins for valuable conversations that informed this work. KMC gratefully acknowledges support from the Marshall Commission and the Cambridge Trust. UB  acknowledges support by ELSA (European Lighthouse on Secure and Safe AI) funded by the European Union under grant agreement No. 101070617; IS acknowledges funding from an NSERC fellowship (567554-2022); KC is supported by the
Hertz Foundation, the Paul and Daisy Soros Fellowship, and an NSF
Graduate Research Fellowship under grant \#1745302.; ML acknowledges funding from MSR; TZX acknowledges support from the OpenPhilanthropy AI Fellowship. VM acknowledges a gift from the Siegel Family Foundation. AW  acknowledges  support  from  a  Turing  AI  Fellowship  under grant  EP/V025279/1,  The  Alan  Turing  Institute,  and the Leverhulme Trust via CFI. TLG acknowledges support from ONR grant N00014-22-1-2813. Views and opinions expressed are however those of the author(s) only and do not necessarily reflect those of the institutions listed above. 
%TC:endignore

\bibliography{main}

\begin{thebibliography}{264}
\providecommand{\natexlab}[1]{#1}
\providecommand{\url}[1]{\texttt{#1}}
\expandafter\ifx\csname urlstyle\endcsname\relax
  \providecommand{\doi}[1]{doi: #1}\else
  \providecommand{\doi}{doi: \begingroup \urlstyle{rm}\Url}\fi

\bibitem[Git()]{GitHubCopilot}
{G}it{H}ub {C}opilot · {Y}our {A}{I} pair programmer.
\newblock \url{https://github.com/features/copilot}.

\bibitem[Cop()]{CopilotMicrosoft}
{C}opilot for {M}icrosoft 365 – {M}icrosoft {A}doption.
\newblock \url{https://adoption.microsoft.com/en-us/copilot/}.

\bibitem[Fedorenko et~al.(2024)Fedorenko, Piantadosi, and Gibson]{fedorenko2024language}
Evelina Fedorenko, Steven~T Piantadosi, and Edward~AF Gibson.
\newblock Language is primarily a tool for communication rather than thought.
\newblock \emph{Nature}, 630\penalty0 (8017):\penalty0 575--586, 2024.

\bibitem[Turing(1950)]{turing1950}
Alan Turing.
\newblock Computing machinery and intelligence.
\newblock \emph{Mind}, 59\penalty0 (236):\penalty0 433, 1950.

\bibitem[Clynes and Kline(1960)]{clynes1960cyborgs}
Manfred~E Clynes and Nathan~S Kline.
\newblock Cyborgs and space.
\newblock \emph{Astronautics}, 14\penalty0 (9):\penalty0 26--27, 1960.

\bibitem[Weizenbaum(1966)]{weizenbaum1966eliza}
Joseph Weizenbaum.
\newblock Eliza—a computer program for the study of natural language communication between man and machine.
\newblock \emph{Communications of the ACM}, 9\penalty0 (1):\penalty0 36--45, 1966.

\bibitem[Shneiderman(2022{\natexlab{a}})]{benHCAI}
Ben Shneiderman.
\newblock {}.
\newblock In \emph{{Human-Centered AI}}. Oxford University Press, 01 2022{\natexlab{a}}.
\newblock ISBN 9780192845290.
\newblock \doi{10.1093/oso/9780192845290.001.0001}.

\bibitem[Bundy(1983)]{Bundy1983TheCM}
Alan Bundy.
\newblock The computer modelling of mathematical reasoning.
\newblock 1983.

\bibitem[Anderson et~al.(1990)Anderson, Boyle, Corbett, and Lewis]{anderson1990cognitive}
John~R Anderson, C~Franklin Boyle, Albert~T Corbett, and Matthew~W Lewis.
\newblock Cognitive modeling and intelligent tutoring.
\newblock 1990.

\bibitem[Bommasani et~al.(2021)Bommasani, Hudson, Adeli, Altman, and et~al]{foundationModels}
Rishi Bommasani, Drew~A. Hudson, Ehsan Adeli, Russ~B. Altman, and Simran~Arora et~al.
\newblock On the opportunities and risks of foundation models.
\newblock \emph{CoRR}, abs/2108.07258, 2021.

\bibitem[Ouyang et~al.(2022)Ouyang, Wu, Jiang, Almeida, and Wainwright]{ouyang2022training}
Long Ouyang, Jeffrey Wu, Xu~Jiang, Diogo Almeida, and Carroll et~al Wainwright.
\newblock Training language models to follow instructions with human feedback.
\newblock \emph{Advances in Neural Information Processing Systems}, 35:\penalty0 27730--27744, 2022.

\bibitem[Christiano et~al.(2017)Christiano, Leike, Brown, Martic, and Legg]{christiano2017deep}
Paul~F Christiano, Jan Leike, Tom Brown, Miljan Martic, and Shane et~al Legg.
\newblock Deep reinforcement learning from human preferences.
\newblock \emph{Advances in neural information processing systems}, 30, 2017.

\bibitem[Lee et~al.(2023{\natexlab{a}})Lee, Liu, Ryu, Watkins, and Du]{lee2023aligning}
Kimin Lee, Hao Liu, Moonkyung Ryu, Olivia Watkins, and Yuqing et~al Du.
\newblock Aligning text-to-image models using human feedback.
\newblock \emph{arXiv preprint arXiv:2302.12192}, 2023{\natexlab{a}}.

\bibitem[Ullman(2023)]{ullman2023large}
Tomer Ullman.
\newblock Large language models fail on trivial alterations to theory-of-mind tasks.
\newblock \emph{arXiv preprint arXiv:2302.08399}, 2023.

\bibitem[Collins et~al.(2024)Collins, Jiang, Frieder, Wong, and Zilka]{collins2024evaluating}
Katherine~M Collins, Albert~Q Jiang, Simon Frieder, Lionel Wong, and Miri et~al Zilka.
\newblock Evaluating language models for mathematics through interactions.
\newblock \emph{Proceedings of the National Academy of Sciences}, 121\penalty0 (24):\penalty0 e2318124121, 2024.

\bibitem[Wong et~al.(2023)Wong, Grand, Lew, Goodman, and Mansinghka]{wong2023word}
Lionel Wong, Gabriel Grand, Alexander~K Lew, Noah~D Goodman, and Vikash K et~al Mansinghka.
\newblock From word models to world models: Translating from natural language to the probabilistic language of thought.
\newblock \emph{arXiv preprint arXiv:2306.12672}, pages arXiv--2306, 2023.

\bibitem[Zhang et~al.(2023{\natexlab{a}})Zhang, Collins, Weller, and Tenenbaum]{zhang2023ai}
Cedegao~E Zhang, Katherine~M Collins, Adrian Weller, and Joshua~B Tenenbaum.
\newblock Ai for mathematics: A cognitive science perspective.
\newblock \emph{arXiv preprint arXiv:2310.13021}, 2023{\natexlab{a}}.

\bibitem[Gweon et~al.(2023)Gweon, Fan, and Kim]{gweon2023socially}
Hyowon Gweon, Judith Fan, and Been Kim.
\newblock Socially intelligent machines that learn from humans and help humans learn.
\newblock \emph{Philosophical Transactions of the Royal Society A}, 381\penalty0 (2251):\penalty0 20220048, 2023.

\bibitem[Mahowald et~al.(2024)Mahowald, Ivanova, Blank, Kanwisher, and Tenenbaum]{mahowald2024dissociating}
Kyle Mahowald, Anna~A Ivanova, Idan~A Blank, Nancy Kanwisher, and Joshua B et~al Tenenbaum.
\newblock Dissociating language and thought in large language models.
\newblock \emph{Trends in Cognitive Sciences}, 2024.

\bibitem[McCoy et~al.(2023)McCoy, Yao, Friedman, Hardy, and Griffiths]{mccoy2023embers}
R~Thomas McCoy, Shunyu Yao, Dan Friedman, Matthew Hardy, and Thomas~L Griffiths.
\newblock Embers of autoregression: Understanding large language models through the problem they are trained to solve.
\newblock \emph{arXiv preprint arXiv:2309.13638}, 2023.

\bibitem[Tenenbaum et~al.(2011)Tenenbaum, Kemp, Griffiths, and Goodman]{tenenbaum2011grow}
Joshua~B Tenenbaum, Charles Kemp, Thomas~L Griffiths, and Noah~D Goodman.
\newblock How to grow a mind: Statistics, structure, and abstraction.
\newblock \emph{science}, 331\penalty0 (6022):\penalty0 1279--1285, 2011.

\bibitem[Griffiths et~al.(2023)Griffiths, Zhu, Grant, and McCoy]{griffiths2023bayes}
Thomas~L. Griffiths, Jian-Qiao Zhu, Erin Grant, and R.~Thomas McCoy.
\newblock Bayes in the age of intelligent machines, 2023.

\bibitem[Sumers et~al.(2023{\natexlab{a}})Sumers, Yao, Narasimhan, and Griffiths]{sumers2023cognitive}
Theodore Sumers, Shunyu Yao, Karthik Narasimhan, and Thomas~L Griffiths.
\newblock Cognitive architectures for language agents.
\newblock \emph{arXiv preprint arXiv:2309.02427}, 2023{\natexlab{a}}.

\bibitem[Binz and Schulz(2023)]{binz2023turning}
Marcel Binz and Eric Schulz.
\newblock Turning large language models into cognitive models.
\newblock \emph{arXiv preprint arXiv:2306.03917}, 2023.

\bibitem[Cusumano-Towner et~al.(2019)Cusumano-Towner, Saad, Lew, and Mansinghka]{cusumano2019gen}
Marco~F Cusumano-Towner, Feras~A Saad, Alexander~K Lew, and Vikash~K Mansinghka.
\newblock Gen: a general-purpose probabilistic programming system with programmable inference.
\newblock In \emph{Proceedings of the 40th acm sigplan conference on programming language design and implementation}, pages 221--236, 2019.

\bibitem[Goodman et~al.(2008{\natexlab{a}})Goodman, Mansinghka, Roy, Bonawitz, and Tenenbaum]{goodman2008church}
Noah~D Goodman, Vikash~K Mansinghka, Daniel Roy, Keith Bonawitz, and Joshua~B Tenenbaum.
\newblock Church: a language for generative models.
\newblock In \emph{Proceedings of the Twenty-Fourth Conference on Uncertainty in Artificial Intelligence}, pages 220--229, 2008{\natexlab{a}}.

\bibitem[Bingham et~al.(2019)Bingham, Chen, Jankowiak, Obermeyer, and Pradhan]{bingham2019pyro}
Eli Bingham, Jonathan~P Chen, Martin Jankowiak, Fritz Obermeyer, and Neeraj et~al Pradhan.
\newblock Pyro: Deep universal probabilistic programming.
\newblock \emph{The Journal of Machine Learning Research}, 20\penalty0 (1):\penalty0 973--978, 2019.

\bibitem[Ge et~al.(2018)Ge, Xu, and Ghahramani]{ge2018turing}
Hong Ge, Kai Xu, and Zoubin Ghahramani.
\newblock Turing: a language for flexible probabilistic inference.
\newblock In \emph{International conference on artificial intelligence and statistics}, pages 1682--1690. PMLR, 2018.

\bibitem[Goodman et~al.(2014)Goodman, Tenenbaum, and Gerstenberg]{goodman2014concepts}
Noah~D Goodman, Joshua~B Tenenbaum, and Tobias Gerstenberg.
\newblock Concepts in a probabilistic language of thought.
\newblock Technical report, Center for Brains, Minds and Machines (CBMM), 2014.

\bibitem[van Opheusden et~al.(2023)van Opheusden, Kuperwajs, Galbiati, Bnaya, and Li]{van2023expertise}
Bas van Opheusden, Ionatan Kuperwajs, Gianni Galbiati, Zahy Bnaya, and Yunqi et~al Li.
\newblock Expertise increases planning depth in human gameplay.
\newblock \emph{Nature}, 618\penalty0 (7967):\penalty0 1000--1005, 2023.

\bibitem[Trinh et~al.(2024)Trinh, Wu, Le, He, and Luong]{trinh2024solving}
Trieu~H Trinh, Yuhuai Wu, Quoc~V Le, He~He, and Thang Luong.
\newblock Solving olympiad geometry without human demonstrations.
\newblock \emph{Nature}, 625\penalty0 (7995):\penalty0 476--482, 2024.

\bibitem[Yao et~al.(2024)Yao, Yu, Zhao, Shafran, and Griffiths]{yao2024tree}
Shunyu Yao, Dian Yu, Jeffrey Zhao, Izhak Shafran, and Tom et~al Griffiths.
\newblock Tree of thoughts: Deliberate problem solving with large language models.
\newblock \emph{Advances in Neural Information Processing Systems}, 36, 2024.

\bibitem[Baker et~al.(2009)Baker, Saxe, and Tenenbaum]{baker2009action}
Chris~L Baker, Rebecca Saxe, and Joshua~B Tenenbaum.
\newblock Action understanding as inverse planning.
\newblock \emph{Cognition}, 113\penalty0 (3):\penalty0 329--349, 2009.

\bibitem[Jara-Ettinger et~al.(2020)Jara-Ettinger, Schulz, and Tenenbaum]{jara2020naive}
Julian Jara-Ettinger, Laura~E Schulz, and Joshua~B Tenenbaum.
\newblock The naive utility calculus as a unified, quantitative framework for action understanding.
\newblock \emph{Cognitive Psychology}, 123:\penalty0 101334, 2020.

\bibitem[Zhi-Xuan et~al.(2024)Zhi-Xuan, Ying, Mansinghka, and Tenenbaum]{zhi2024pragmatic}
Tan Zhi-Xuan, Lance Ying, Vikash Mansinghka, and Joshua~B Tenenbaum.
\newblock Pragmatic instruction following and goal assistance via cooperative language-guided inverse planning.
\newblock In \emph{Proceedings of the 23rd International Conference on Autonomous Agents and Multiagent Systems}, pages 2094--2103, 2024.

\bibitem[Lake et~al.(2017)Lake, Ullman, Tenenbaum, and Gershman]{lake_ullman_tenenbaum_gershman_2017}
Brenden~M. Lake, Tomer~D. Ullman, Joshua~B. Tenenbaum, and Samuel~J. Gershman.
\newblock Building machines that learn and think like people.
\newblock \emph{Behavioral and Brain Sciences}, 40:\penalty0 e253, 2017.
\newblock \doi{10.1017/S0140525X16001837}.

\bibitem[Chu and Schulz(2020)]{chu2020play}
Junyi Chu and Laura~E Schulz.
\newblock Play, curiosity, and cognition.
\newblock \emph{Annual Review of Developmental Psychology}, 2\penalty0 (1):\penalty0 317--343, 2020.

\bibitem[Yanai and Lercher(2024)]{yanai2024takes}
Itai Yanai and Martin~J Lercher.
\newblock It takes two to think.
\newblock \emph{Nature Biotechnology}, pages 1--2, 2024.

\bibitem[Holyoak and Morrison(2005)]{holyoak2005cambridge}
Keith~J Holyoak and Robert~G Morrison.
\newblock \emph{The Cambridge handbook of thinking and reasoning}.
\newblock Cambridge University Press, 2005.

\bibitem[Holyoak and Morrison(2012)]{holyoak2012oxford}
Keith~J Holyoak and Robert~G Morrison.
\newblock \emph{The Oxford handbook of thinking and reasoning}.
\newblock Oxford University Press, 2012.

\bibitem[Ko and Myers(2004)]{ko2004designing}
Amy~J Ko and Brad~A Myers.
\newblock Designing the whyline: a debugging interface for asking questions about program behavior.
\newblock In \emph{Proceedings of the SIGCHI conference on Human factors in computing systems}, pages 151--158, 2004.

\bibitem[Ko et~al.(2011)Ko, Abraham, Beckwith, Blackwell, and Burnett]{ko2011state}
Amy~J Ko, Robin Abraham, Laura Beckwith, Alan Blackwell, and Margaret et~al Burnett.
\newblock The state of the art in end-user software engineering.
\newblock \emph{ACM Computing Surveys (CSUR)}, 43\penalty0 (3):\penalty0 1--44, 2011.

\bibitem[Muggleton and De~Raedt(1994)]{muggleton1994inductive}
Stephen Muggleton and Luc De~Raedt.
\newblock Inductive logic programming: Theory and methods.
\newblock \emph{The Journal of Logic Programming}, 19:\penalty0 629--679, 1994.

\bibitem[Anderson and Reiser(1985)]{anderson1985lisp}
John~R Anderson and Brian~J Reiser.
\newblock The lisp tutor.
\newblock \emph{Byte}, 10\penalty0 (4):\penalty0 159--175, 1985.

\bibitem[Anderson et~al.(1995)Anderson, Corbett, Koedinger, and Pelletier]{anderson1995cognitive}
John~R Anderson, Albert~T Corbett, Kenneth~R Koedinger, and Ray Pelletier.
\newblock Cognitive tutors: Lessons learned.
\newblock \emph{The journal of the learning sciences}, 4\penalty0 (2):\penalty0 167--207, 1995.

\bibitem[Imai(2022)]{imai2022github}
Saki Imai.
\newblock Is github copilot a substitute for human pair-programming? an empirical study.
\newblock In \emph{Proceedings of the ACM/IEEE 44th International Conference on Software Engineering: Companion Proceedings}, pages 319--321, 2022.

\bibitem[Nguyen and Nadi(2022)]{nguyen2022empirical}
Nhan Nguyen and Sarah Nadi.
\newblock An empirical evaluation of github copilot's code suggestions.
\newblock In \emph{Proceedings of the 19th International Conference on Mining Software Repositories}, pages 1--5, 2022.

\bibitem[Wermelinger(2023)]{wermelinger2023using}
Michel Wermelinger.
\newblock Using github copilot to solve simple programming problems.
\newblock In \emph{Proceedings of the 54th ACM Technical Symposium on Computer Science Education V. 1}, pages 172--178, 2023.

\bibitem[Barke et~al.(2023)Barke, James, and Polikarpova]{barke2023grounded}
Shraddha Barke, Michael~B James, and Nadia Polikarpova.
\newblock Grounded copilot: How programmers interact with code-generating models.
\newblock \emph{Proceedings of the ACM on Programming Languages}, 7\penalty0 (OOPSLA1):\penalty0 85--111, 2023.

\bibitem[Dakhel et~al.(2023)Dakhel, Majdinasab, Nikanjam, Khomh, and Desmarais]{dakhel2023github}
Arghavan~Moradi Dakhel, Vahid Majdinasab, Amin Nikanjam, Foutse Khomh, and Michel C et~al Desmarais.
\newblock Github copilot ai pair programmer: Asset or liability?
\newblock \emph{Journal of Systems and Software}, 203:\penalty0 111734, 2023.

\bibitem[Fisac et~al.(2020{\natexlab{a}})Fisac, Gates, Hamrick, Liu, and Hadfield-Menell]{fisac2020pragmatic}
Jaime~F Fisac, Monica~A Gates, Jessica~B Hamrick, Chang Liu, and Dylan et~al Hadfield-Menell.
\newblock Pragmatic-pedagogic value alignment.
\newblock In \emph{Robotics Research: The 18th International Symposium ISRR}, pages 49--57. Springer, 2020{\natexlab{a}}.

\bibitem[Ranz et~al.(2017)Ranz, Hummel, and Sihn]{ranz2017capability}
Fabian Ranz, Vera Hummel, and Wilfried Sihn.
\newblock Capability-based task allocation in human-robot collaboration.
\newblock \emph{Procedia Manufacturing}, 9:\penalty0 182--189, 2017.

\bibitem[Casper and Murphy(2003)]{casper2003human}
Jennifer Casper and Robin~R. Murphy.
\newblock Human-robot interactions during the robot-assisted urban search and rescue response at the world trade center.
\newblock \emph{IEEE Transactions on Systems, Man, and Cybernetics, Part B (Cybernetics)}, 33\penalty0 (3):\penalty0 367--385, 2003.

\bibitem[Shridhar et~al.(2020)Shridhar, Thomason, Gordon, Bisk, and Han]{shridhar2020alfred}
Mohit Shridhar, Jesse Thomason, Daniel Gordon, Yonatan Bisk, and Winson et~al Han.
\newblock Alfred: A benchmark for interpreting grounded instructions for everyday tasks.
\newblock In \emph{Proceedings of the IEEE/CVF conference on computer vision and pattern recognition}, pages 10740--10749, 2020.

\bibitem[Ahn et~al.(2022)Ahn, Brohan, Brown, Chebotar, and Cortes]{ahn2022can}
Michael Ahn, Anthony Brohan, Noah Brown, Yevgen Chebotar, and Omar et~al Cortes.
\newblock Do as i can, not as i say: Grounding language in robotic affordances.
\newblock \emph{arXiv preprint arXiv:2204.01691}, 2022.
\newblock Last Accessed: {2024–07-07}.

\bibitem[Raad et~al.(2024)Raad, Ahuja, Barros, Besse, and Bolt]{raad2024scaling}
Maria~Abi Raad, Arun Ahuja, Catarina Barros, Frederic Besse, and Andrew et~al Bolt.
\newblock Scaling instructable agents across many simulated worlds.
\newblock \emph{arXiv preprint arXiv:2404.10179}, 2024.

\bibitem[Valmeekam et~al.(2022)Valmeekam, Olmo, Sreedharan, and Kambhampati]{valmeekam2022large}
Karthik Valmeekam, Alberto Olmo, Sarath Sreedharan, and Subbarao Kambhampati.
\newblock Large language models still can't plan (a benchmark for llms on planning and reasoning about change).
\newblock \emph{arXiv preprint arXiv:2206.10498}, 2022.

\bibitem[Momennejad et~al.(2024)Momennejad, Hasanbeig, Vieira~Frujeri, Sharma, and Jojic]{momennejad2024evaluating}
Ida Momennejad, Hosein Hasanbeig, Felipe Vieira~Frujeri, Hiteshi Sharma, and Nebojsa et~al Jojic.
\newblock Evaluating cognitive maps and planning in large language models with cogeval.
\newblock \emph{Advances in Neural Information Processing Systems}, 36, 2024.

\bibitem[Goodman and Frank(2016)]{goodman2016pragmatic}
Noah~D Goodman and Michael~C Frank.
\newblock Pragmatic language interpretation as probabilistic inference.
\newblock \emph{Trends in cognitive sciences}, 20\penalty0 (11):\penalty0 818--829, 2016.

\bibitem[Sumers et~al.(2023{\natexlab{b}})Sumers, Ho, Griffiths, and Hawkins]{sumers2023reconciling}
Theodore~R Sumers, Mark~K Ho, Thomas~L Griffiths, and Robert~D Hawkins.
\newblock Reconciling truthfulness and relevance as epistemic and decision-theoretic utility.
\newblock \emph{Psychological Review}, 2023{\natexlab{b}}.

\bibitem[Jeon et~al.(2020)Jeon, Milli, and Dragan]{jeon2020reward}
Hong~Jun Jeon, Smitha Milli, and Anca Dragan.
\newblock Reward-rational (implicit) choice: A unifying formalism for reward learning.
\newblock \emph{Advances in Neural Information Processing Systems}, 33:\penalty0 4415--4426, 2020.

\bibitem[Kollar et~al.(2013)Kollar, Tellex, Walter, Huang, and Bachrach]{kollar2013generalized}
Thomas Kollar, Stefanie Tellex, Matthew~R Walter, Albert Huang, and Abraham et~al Bachrach.
\newblock Generalized grounding graphs: A probabilistic framework for understanding grounded language.
\newblock \emph{Journal of Artificial Intelligence Research}, pages 1--35, 2013.

\bibitem[Bratman(2013)]{bratman2013shared}
Michael~E Bratman.
\newblock \emph{Shared agency: A planning theory of acting together}.
\newblock Oxford University Press, 2013.

\bibitem[Stacy et~al.(2021)Stacy, Li, Zhao, Yun, and Zhao]{stacy2021modeling}
Stephanie Stacy, Chenfei Li, Minglu Zhao, Yiling Yun, and Qingyi et~al Zhao.
\newblock Modeling communication to coordinate perspectives in cooperation.
\newblock In \emph{Proceedings of the Annual Meeting of the Cognitive Science Society}, volume~43, 2021.

\bibitem[Wu et~al.(2021)Wu, Wang, Evans, Tenenbaum, and Parkes]{wu2021too}
Sarah~A Wu, Rose~E Wang, James~A Evans, Joshua~B Tenenbaum, and David C et~al Parkes.
\newblock Too many cooks: Bayesian inference for coordinating multi-agent collaboration.
\newblock \emph{Topics in Cognitive Science}, 13\penalty0 (2):\penalty0 414--432, 2021.

\bibitem[Reddy et~al.(2018)Reddy, Dragan, and Levine]{reddy2018you}
Sid Reddy, Anca Dragan, and Sergey Levine.
\newblock Where do you think you're going?: Inferring beliefs about dynamics from behavior.
\newblock \emph{Advances in Neural Information Processing Systems}, 31, 2018.

\bibitem[Alanqary et~al.(2021)Alanqary, Lin, Le, Zhi-Xuan, and Mansinghka]{alanqary2021modeling}
Arwa Alanqary, Gloria~Z Lin, Joie Le, Tan Zhi-Xuan, and Vikash K et~al Mansinghka.
\newblock Modeling the mistakes of boundedly rational agents within a bayesian theory of mind.
\newblock \emph{arXiv preprint arXiv:2106.13249}, 2021.

\bibitem[Dragan et~al.(2013)Dragan, Lee, and Srinivasa]{dragan2013legibility}
Anca~D Dragan, Kenton~CT Lee, and Siddhartha~S Srinivasa.
\newblock Legibility and predictability of robot motion.
\newblock In \emph{2013 8th ACM/IEEE International Conference on Human-Robot Interaction (HRI)}, pages 301--308. IEEE, 2013.

\bibitem[Miura and Zilberstein(2021)]{miura2021unifying}
Shuwa Miura and Shlomo Zilberstein.
\newblock A unifying framework for observer-aware planning and its complexity.
\newblock In \emph{Uncertainty in Artificial Intelligence}, pages 610--620. PMLR, 2021.

\bibitem[Flower and Hayes(1981)]{flower1981cognitive}
Linda Flower and John~R. Hayes.
\newblock A cognitive process theory of writing.
\newblock \emph{College Composition and Communication}, 32\penalty0 (4):\penalty0 365--387, 1981.
\newblock ISSN 0010096X.

\bibitem[Hayes(2012)]{hayes2012modeling}
John~R Hayes.
\newblock Modeling and remodeling writing.
\newblock \emph{Written communication}, 29\penalty0 (3):\penalty0 369--388, 2012.

\bibitem[Lee et~al.(2022)Lee, Liang, and Yang]{lee2022coauthor}
Mina Lee, Percy Liang, and Qian Yang.
\newblock Coauthor: Designing a human-ai collaborative writing dataset for exploring language model capabilities.
\newblock In \emph{Proceedings of the 2022 CHI Conference on Human Factors in Computing Systems}, pages 1--19, 2022.

\bibitem[Lee et~al.(2024)Lee, Gero, Chung, Shum, and Raheja]{lee2024design}
Mina Lee, Katy~Ilonka Gero, John Joon~Young Chung, Simon~Buckingham Shum, and Vipul et~al Raheja.
\newblock A design space for intelligent and interactive writing assistants.
\newblock \emph{CHI}, 2024.

\bibitem[Ippolito et~al.(2022)Ippolito, Yuan, Coenen, and Burnam]{ippolito2022creative}
Daphne Ippolito, Ann Yuan, Andy Coenen, and Sehmon Burnam.
\newblock Creative writing with an ai-powered writing assistant: Perspectives from professional writers, 2022.

\bibitem[Gero et~al.(2022)Gero, Liu, and Chilton]{gero2022sparks}
Katy~Ilonka Gero, Vivian Liu, and Lydia Chilton.
\newblock Sparks: Inspiration for science writing using language models.
\newblock In \emph{Proceedings of the 2022 ACM Designing Interactive Systems Conference}, pages 1002--1019, 2022.

\bibitem[Gero et~al.(2023)Gero, Long, and Chilton]{gero2023social}
Katy~Ilonka Gero, Tao Long, and Lydia~B Chilton.
\newblock Social dynamics of ai support in creative writing.
\newblock In \emph{Proceedings of the 2023 CHI Conference on Human Factors in Computing Systems}, pages 1--15, 2023.

\bibitem[Dell'Acqua et~al.(2023)Dell'Acqua, McFowland, Mollick, Lifshitz-Assaf, and Kellogg]{dell2023navigating}
Fabrizio Dell'Acqua, Edward McFowland, Ethan~R Mollick, Hila Lifshitz-Assaf, and Katherine et~al Kellogg.
\newblock Navigating the jagged technological frontier: Field experimental evidence of the effects of ai on knowledge worker productivity and quality.
\newblock \emph{Harvard Business School Technology \& Operations Mgt. Unit Working Paper}, \penalty0 (24-013), 2023.

\bibitem[Porter et~al.(2023)Porter, Boyd, Skandari, and Laiteerapong]{porter2023revisiting}
Justin Porter, Cynthia Boyd, M~Reza Skandari, and Neda Laiteerapong.
\newblock Revisiting the time needed to provide adult primary care.
\newblock \emph{Journal of general internal medicine}, 38\penalty0 (1):\penalty0 147--155, 2023.

\bibitem[Dewa et~al.(2017)Dewa, Loong, Bonato, and Trojanowski]{dewa2017relationship}
Carolyn~S Dewa, Desmond Loong, Sarah Bonato, and Lucy Trojanowski.
\newblock The relationship between physician burnout and quality of healthcare in terms of safety and acceptability: a systematic review.
\newblock \emph{BMJ open}, 7\penalty0 (6):\penalty0 e015141, 2017.

\bibitem[Chowdhery et~al.(2022)Chowdhery, Narang, Devlin, Bosma, and Mishra]{chowdhery2022palm}
Aakanksha Chowdhery, Sharan Narang, Jacob Devlin, Maarten Bosma, and Gaurav et~al Mishra.
\newblock Palm: Scaling language modeling with pathways.
\newblock \emph{arXiv preprint arXiv:2204.02311}, 2022.

\bibitem[Singhal et~al.(2023)Singhal, Azizi, Tu, Mahdavi, and Wei]{singhal2023large}
Karan Singhal, Shekoofeh Azizi, Tao Tu, S~Sara Mahdavi, and Jason et~al Wei.
\newblock Large language models encode clinical knowledge.
\newblock \emph{Nature}, 620\penalty0 (7972):\penalty0 172--180, 2023.

\bibitem[Topol(2019)]{topol2019high}
Eric~J Topol.
\newblock High-performance medicine: the convergence of human and artificial intelligence.
\newblock \emph{Nature medicine}, 25\penalty0 (1):\penalty0 44--56, 2019.

\bibitem[Tu et~al.(2024)Tu, Palepu, Schaekermann, Saab, and et~al]{tu2024conversational}
Tao Tu, Anil Palepu, Mike Schaekermann, Khaled Saab, and Jan~Freyberg et~al.
\newblock Towards conversational diagnostic ai, 2024.

\bibitem[Ayers et~al.(2023)Ayers, Poliak, Dredze, Leas, and Zhu]{chatbotMed}
John~W. Ayers, Adam Poliak, Mark Dredze, Eric~C. Leas, and Zechariah et~al Zhu.
\newblock {Comparing Physician and Artificial Intelligence Chatbot Responses to Patient Questions Posted to a Public Social Media Forum}.
\newblock \emph{JAMA Internal Medicine}, 04 2023.
\newblock ISSN 2168-6106.
\newblock \doi{10.1001/jamainternmed.2023.1838}.

\bibitem[Vellido(2020)]{vellido2020importance}
Alfredo Vellido.
\newblock The importance of interpretability and visualization in machine learning for applications in medicine and health care.
\newblock \emph{Neural computing and applications}, 32\penalty0 (24):\penalty0 18069--18083, 2020.

\bibitem[Rajpurkar et~al.(2022)Rajpurkar, Chen, Banerjee, and Topol]{rajpurkar2022ai}
Pranav Rajpurkar, Emma Chen, Oishi Banerjee, and Eric~J Topol.
\newblock Ai in health and medicine.
\newblock \emph{Nature medicine}, 28\penalty0 (1):\penalty0 31--38, 2022.

\bibitem[Ghassemi et~al.(2020)Ghassemi, Naumann, Schulam, Beam, and Chen]{ghassemi2020review}
Marzyeh Ghassemi, Tristan Naumann, Peter Schulam, Andrew~L Beam, and Irene Y et~al Chen.
\newblock A review of challenges and opportunities in machine learning for health.
\newblock \emph{AMIA Summits on Translational Science Proceedings}, 2020:\penalty0 191, 2020.

\bibitem[Daneshjou et~al.(2021)Daneshjou, Smith, Sun, Rotemberg, and Zou]{daneshjou2021lack}
Roxana Daneshjou, Mary~P Smith, Mary~D Sun, Veronica Rotemberg, and James Zou.
\newblock Lack of transparency and potential bias in artificial intelligence data sets and algorithms: a scoping review.
\newblock \emph{JAMA dermatology}, 157\penalty0 (11):\penalty0 1362--1369, 2021.

\bibitem[Cabitza and Zeitoun(2019)]{cabitza2019proof}
Federico Cabitza and Jean-David Zeitoun.
\newblock The proof of the pudding: in praise of a culture of real-world validation for medical artificial intelligence.
\newblock \emph{Annals of translational medicine}, 7\penalty0 (8), 2019.

\bibitem[Puig et~al.(2020)Puig, Shu, Li, Wang, and Liao]{puig2020watch}
Xavier Puig, Tianmin Shu, Shuang Li, Zilin Wang, and Yuan-Hong et~al Liao.
\newblock Watch-and-help: A challenge for social perception and human-ai collaboration.
\newblock \emph{arXiv preprint arXiv:2010.09890}, 2020.

\bibitem[Chandra et~al.(2024{\natexlab{a}})Chandra, Chen, Li, Ragan-Kelley, and Tenenbaum]{chandra2024inferring}
Kartik Chandra, Tony Chen, Tzu-Mao Li, Jonathan Ragan-Kelley, and Josh Tenenbaum.
\newblock Inferring the future by imagining the past.
\newblock \emph{Advances in Neural Information Processing Systems}, 36, 2024{\natexlab{a}}.

\bibitem[Fisac et~al.(2020{\natexlab{b}})Fisac, Liu, Hamrick, Sastry, and Hedrick]{fisac2020generating}
Jaime~F Fisac, Chang Liu, Jessica~B Hamrick, Shankar Sastry, and J~Karl et~al Hedrick.
\newblock Generating plans that predict themselves.
\newblock In \emph{Algorithmic Foundations of Robotics XII: Proceedings of the Twelfth Workshop on the Algorithmic Foundations of Robotics}, pages 144--159. Springer, 2020{\natexlab{b}}.

\bibitem[Grice(1975)]{grice1975logic}
Herbert~P Grice.
\newblock Logic and conversation.
\newblock In \emph{Speech acts}, pages 41--58. Brill, 1975.

\bibitem[Doshi-Velez and Kim(2017)]{doshi2017towards}
Finale Doshi-Velez and Been Kim.
\newblock Towards a rigorous science of interpretable machine learning.
\newblock \emph{arXiv preprint arXiv:1702.08608}, 2017.

\bibitem[Miller(2019)]{miller2019explanation}
Tim Miller.
\newblock Explanation in artificial intelligence: Insights from the social sciences.
\newblock \emph{Artificial intelligence}, 267:\penalty0 1--38, 2019.

\bibitem[Smith(2019)]{smith2019promise}
Brian~Cantwell Smith.
\newblock \emph{The promise of artificial intelligence: reckoning and judgment}.
\newblock Mit Press, 2019.

\bibitem[Sucholutsky and Griffiths(2023)]{sucholutsky2023alignment}
Ilia Sucholutsky and Thomas~L Griffiths.
\newblock Alignment with human representations supports robust few-shot learning.
\newblock \emph{NeurIPS}, 2023.

\bibitem[Sucholutsky et~al.(2023)Sucholutsky, Muttenthaler, Weller, Peng, and et~al]{sucholutsky2023getting}
Ilia Sucholutsky, Lukas Muttenthaler, Adrian Weller, Andi Peng, and Andreea~Bobu et~al.
\newblock Getting aligned on representational alignment, 2023.

\bibitem[Battaglia et~al.(2013)Battaglia, Hamrick, and Tenenbaum]{battaglia2013simulation}
Peter~W Battaglia, Jessica~B Hamrick, and Joshua~B Tenenbaum.
\newblock Simulation as an engine of physical scene understanding.
\newblock \emph{Proceedings of the National Academy of Sciences}, 110\penalty0 (45):\penalty0 18327--18332, 2013.

\bibitem[Roncone et~al.(2017)Roncone, Mangin, and Scassellati]{roncone2017transparent}
Alessandro Roncone, Olivier Mangin, and Brian Scassellati.
\newblock Transparent role assignment and task allocation in human robot collaboration.
\newblock In \emph{2017 IEEE International Conference on Robotics and Automation (ICRA)}, pages 1014--1021. IEEE, 2017.

\bibitem[Carroll et~al.(2019)Carroll, Shah, Ho, Griffiths, and Seshia]{carroll2019utility}
Micah Carroll, Rohin Shah, Mark~K Ho, Tom Griffiths, and Sanjit et~al Seshia.
\newblock On the utility of learning about humans for human-ai coordination.
\newblock \emph{Advances in neural information processing systems}, 32, 2019.

\bibitem[Macindoe et~al.(2012)Macindoe, Kaelbling, and Lozano-P{\'e}rez]{macindoe2012pomcop}
Owen Macindoe, Leslie~Pack Kaelbling, and Tom{\'a}s Lozano-P{\'e}rez.
\newblock Pomcop: Belief space planning for sidekicks in cooperative games.
\newblock In \emph{Proceedings of the AAAI Conference on Artificial Intelligence and Interactive Digital Entertainment}, volume~8, pages 38--43, 2012.

\bibitem[Lin et~al.(2022)Lin, Fried, Klein, and Dragan]{lin2022inferring}
Jessy Lin, Daniel Fried, Dan Klein, and Anca Dragan.
\newblock Inferring rewards from language in context.
\newblock \emph{arXiv preprint arXiv:2204.02515}, 2022.

\bibitem[Keuning et~al.(2018)Keuning, Jeuring, and Heeren]{keuning2018systematic}
Hieke Keuning, Johan Jeuring, and Bastiaan Heeren.
\newblock A systematic literature review of automated feedback generation for programming exercises.
\newblock \emph{ACM Transactions on Computing Education (TOCE)}, 19\penalty0 (1):\penalty0 1--43, 2018.

\bibitem[Sarsa et~al.(2022)Sarsa, Denny, Hellas, and Leinonen]{sarsa2022automatic}
Sami Sarsa, Paul Denny, Arto Hellas, and Juho Leinonen.
\newblock Automatic generation of programming exercises and code explanations using large language models.
\newblock In \emph{Proceedings of the 2022 ACM Conference on International Computing Education Research-Volume 1}, pages 27--43, 2022.

\bibitem[Chandra et~al.(2024{\natexlab{b}})Chandra, Li, Nigam, Tenenbaum, and Ragan-Kelley]{chandra2024watchat}
Kartik Chandra, Tzu-Mao Li, Rachit Nigam, Joshua Tenenbaum, and Jonathan Ragan-Kelley.
\newblock Watchat: Explaining perplexing programs by debugging mental models, 2024{\natexlab{b}}.

\bibitem[Head et~al.(2015)Head, Appachu, Hearst, and Hartmann]{head2015tutorons}
Andrew Head, Codanda Appachu, Marti~A Hearst, and Bj{\"o}rn Hartmann.
\newblock Tutorons: Generating context-relevant, on-demand explanations and demonstrations of online code.
\newblock In \emph{2015 IEEE Symposium on Visual Languages and Human-Centric Computing (VL/HCC)}, pages 3--12. IEEE, 2015.

\bibitem[Rafferty et~al.(2020)Rafferty, Jansen, and Griffiths]{rafferty2020assessing}
Anna~N Rafferty, Rachel~A Jansen, and Thomas~L Griffiths.
\newblock Assessing mathematics misunderstandings via bayesian inverse planning.
\newblock \emph{Cognitive science}, 44\penalty0 (10):\penalty0 e12900, 2020.

\bibitem[Poesia and Goodman(2023)]{poesia2023peano}
Gabriel Poesia and Noah~D Goodman.
\newblock Peano: learning formal mathematical reasoning.
\newblock \emph{Philosophical Transactions of the Royal Society A}, 381\penalty0 (2251):\penalty0 20220044, 2023.

\bibitem[Slonim et~al.(2021)Slonim, Bilu, Alzate, Bar-Haim, and Bogin]{slonim2021autonomous}
Noam Slonim, Yonatan Bilu, Carlos Alzate, Roy Bar-Haim, and Ben et~al Bogin.
\newblock An autonomous debating system.
\newblock \emph{Nature}, 591\penalty0 (7850):\penalty0 379--384, 2021.

\bibitem[Jarrett et~al.(2023)Jarrett, Pislar, Bakker, Tessler, and Koster]{jarrett2023language}
Daniel Jarrett, Miruna Pislar, Michiel~A Bakker, Michael~Henry Tessler, and Raphael et~al Koster.
\newblock Language agents as digital representatives in collective decision-making.
\newblock In \emph{NeurIPS 2023 Foundation Models for Decision Making Workshop}, 2023.

\bibitem[Du et~al.(2023)Du, Li, Torralba, Tenenbaum, and Mordatch]{du2023improving}
Yilun Du, Shuang Li, Antonio Torralba, Joshua~B Tenenbaum, and Igor Mordatch.
\newblock Improving factuality and reasoning in language models through multiagent debate.
\newblock \emph{arXiv preprint arXiv:2305.14325}, 2023.

\bibitem[Bakker et~al.(2022)Bakker, Chadwick, Sheahan, Tessler, and Campbell-Gillingham]{bakker2022fine}
Michiel Bakker, Martin Chadwick, Hannah Sheahan, Michael Tessler, and Lucy et~al Campbell-Gillingham.
\newblock Fine-tuning language models to find agreement among humans with diverse preferences.
\newblock \emph{Advances in Neural Information Processing Systems}, 35:\penalty0 38176--38189, 2022.

\bibitem[Small et~al.(2021)Small, Bjorkegren, Erkkil{\"a}, Shaw, and Megill]{small2021polis}
Christopher Small, Michael Bjorkegren, Timo Erkkil{\"a}, Lynette Shaw, and Colin Megill.
\newblock Polis: Scaling deliberation by mapping high dimensional opinion spaces.
\newblock \emph{Recerca: revista de pensament i an{\`a}lisi}, 26\penalty0 (2), 2021.

\bibitem[Lew et~al.(2021)Lew, Agrawal, Sontag, and Mansinghka]{lew2021pclean}
Alexander Lew, Monica Agrawal, David Sontag, and Vikash Mansinghka.
\newblock Pclean: Bayesian data cleaning at scale with domain-specific probabilistic programming.
\newblock In \emph{International Conference on Artificial Intelligence and Statistics}, pages 1927--1935. PMLR, 2021.

\bibitem[Saad et~al.(2019)Saad, Cusumano-Towner, Schaechtle, Rinard, and Mansinghka]{saad2019bayesian}
Feras~A Saad, Marco~F Cusumano-Towner, Ulrich Schaechtle, Martin~C Rinard, and Vikash~K Mansinghka.
\newblock Bayesian synthesis of probabilistic programs for automatic data modeling.
\newblock \emph{Proceedings of the ACM on Programming Languages}, 3\penalty0 (POPL):\penalty0 1--32, 2019.

\bibitem[Huot et~al.(2024)Huot, Ghavami, Lew, Schaechtle, and Freer]{huot2024gensql}
Mathieu Huot, Matin Ghavami, Alexander~K Lew, Ulrich Schaechtle, and Cameron E et~al Freer.
\newblock Gensql: A probabilistic programming system for querying generative models of database tables.
\newblock \emph{Proceedings of the ACM on Programming Languages}, 8\penalty0 (PLDI):\penalty0 790--815, 2024.

\bibitem[Li et~al.(2024)Li, Fox, and Goodman]{li2024automated}
Michael~Y Li, Emily~B Fox, and Noah~D Goodman.
\newblock Automated statistical model discovery with language models.
\newblock \emph{arXiv preprint arXiv:2402.17879}, 2024.

\bibitem[Steinruecken et~al.(2019)Steinruecken, Smith, Janz, Lloyd, and Ghahramani]{steinruecken2019automatic}
Christian Steinruecken, Emma Smith, David Janz, James Lloyd, and Zoubin Ghahramani.
\newblock The automatic statistician.
\newblock \emph{Automated machine learning: Methods, systems, challenges}, pages 161--173, 2019.

\bibitem[Davies et~al.(2021)Davies, Veli{\v{c}}kovi{\'c}, Buesing, Blackwell, and Zheng]{davies2021advancing}
Alex Davies, Petar Veli{\v{c}}kovi{\'c}, Lars Buesing, Sam Blackwell, and Daniel et~al Zheng.
\newblock Advancing mathematics by guiding human intuition with ai.
\newblock \emph{Nature}, 600\penalty0 (7887):\penalty0 70--74, 2021.

\bibitem[Cranmer et~al.(2020)Cranmer, Sanchez~Gonzalez, Battaglia, Xu, and Cranmer]{cranmer2020discovering}
Miles Cranmer, Alvaro Sanchez~Gonzalez, Peter Battaglia, Rui Xu, and Kyle et~al Cranmer.
\newblock Discovering symbolic models from deep learning with inductive biases.
\newblock \emph{Advances in neural information processing systems}, 33:\penalty0 17429--17442, 2020.

\bibitem[Romera-Paredes et~al.(2024)Romera-Paredes, Barekatain, Novikov, Balog, and Kumar]{romera2024mathematical}
Bernardino Romera-Paredes, Mohammadamin Barekatain, Alexander Novikov, Matej Balog, and M~Pawan et~al Kumar.
\newblock Mathematical discoveries from program search with large language models.
\newblock \emph{Nature}, 625\penalty0 (7995):\penalty0 468--475, 2024.

\bibitem[Ashkinaze et~al.(2024)Ashkinaze, Mendelsohn, Qiwei, Budak, and Gilbert]{ashkinaze2024ai}
Joshua Ashkinaze, Julia Mendelsohn, Li~Qiwei, Ceren Budak, and Eric Gilbert.
\newblock How ai ideas affect the creativity, diversity, and evolution of human ideas: Evidence from a large, dynamic experiment, 2024.

\bibitem[Suri et~al.(2024)Suri, Counts, Wang, Chen, and Wan]{suri2024the}
Siddharth Suri, Scott Counts, Leijie Wang, Chacha Chen, and Mengting et~al Wan.
\newblock The use of generative search engines for knowledge work and complex tasks.
\newblock Technical Report MSR-TR-2024-9, Microsoft, March 2024.

\bibitem[Vartiainen and Tedre(2023)]{vartiainen2023using}
Henriikka Vartiainen and Matti Tedre.
\newblock Using artificial intelligence in craft education: crafting with text-to-image generative models.
\newblock \emph{Digital Creativity}, 34\penalty0 (1):\penalty0 1--21, 2023.

\bibitem[Gafni et~al.(2022)Gafni, Polyak, Ashual, Sheynin, and Parikh]{gafni2022make}
Oran Gafni, Adam Polyak, Oron Ashual, Shelly Sheynin, and Devi et~al Parikh.
\newblock Make-a-scene: Scene-based text-to-image generation with human priors.
\newblock In \emph{European Conference on Computer Vision}, pages 89--106. Springer, 2022.

\bibitem[Fan et~al.(2019)Fan, Dinculescu, and Ha]{fan2019collabdraw}
Judith~E Fan, Monica Dinculescu, and David Ha.
\newblock Collabdraw: an environment for collaborative sketching with an artificial agent.
\newblock In \emph{Proceedings of the 2019 Conference on Creativity and Cognition}, pages 556--561, 2019.

\bibitem[Ge et~al.(2020)Ge, Goswami, Zitnick, and Parikh]{ge2020creative}
Songwei Ge, Vedanuj Goswami, C~Lawrence Zitnick, and Devi Parikh.
\newblock Creative sketch generation.
\newblock \emph{arXiv preprint arXiv:2011.10039}, 2020.

\bibitem[Dvoro{\v{z}}{\v{n}}{\'a}k et~al.(2020)Dvoro{\v{z}}{\v{n}}{\'a}k, S{\`y}kora, Curtis, Curless, and Sorkine-Hornung]{dvorovzvnak2020monster}
Marek Dvoro{\v{z}}{\v{n}}{\'a}k, Daniel S{\`y}kora, Cassidy Curtis, Brian Curless, and Olga et~al Sorkine-Hornung.
\newblock Monster mash: a single-view approach to casual 3d modeling and animation.
\newblock \emph{ACM Transactions on Graphics (ToG)}, 39\penalty0 (6):\penalty0 1--12, 2020.

\bibitem[Chater and Oaksford(2008)]{chater2008probabilistic}
Nick Chater and Mike Oaksford, editors.
\newblock \emph{The Probabilistic Mind: Prospects for Bayesian Cognitive Science}.
\newblock Oxford University Press, Oxford, UK, 2008.

\bibitem[Griffiths et~al.(2008)Griffiths, Kemp, and Tenenbaum]{griffiths2008bayesian}
Thomas~L. Griffiths, Charles Kemp, and Joshua~B. Tenenbaum.
\newblock Bayesian models of cognition.
\newblock In Ron Sun, editor, \emph{The Cambridge handbook of computational psychology}, chapter~3, pages 59--100. Cambridge University Press, 2008.

\bibitem[Spelke(2000)]{spelke2000core}
Elizabeth~S Spelke.
\newblock Core knowledge.
\newblock \emph{American psychologist}, 55\penalty0 (11):\penalty0 1233, 2000.

\bibitem[Piantadosi(2021)]{piantadosi2021computational}
Steven~T Piantadosi.
\newblock The computational origin of representation.
\newblock \emph{Minds and machines}, 31\penalty0 (1):\penalty0 1--58, 2021.

\bibitem[Quilty-Dunn et~al.(2023)Quilty-Dunn, Porot, and Mandelbaum]{quilty2023best}
Jake Quilty-Dunn, Nicolas Porot, and Eric Mandelbaum.
\newblock The best game in town: The reemergence of the language-of-thought hypothesis across the cognitive sciences.
\newblock \emph{Behavioral and Brain Sciences}, 46:\penalty0 e261, 2023.

\bibitem[Griffiths et~al.(2003)Griffiths, Jordan, Tenenbaum, and Blei]{griffiths2003hierarchical}
Thomas Griffiths, Michael Jordan, Joshua Tenenbaum, and David Blei.
\newblock Hierarchical topic models and the nested chinese restaurant process.
\newblock \emph{Advances in neural information processing systems}, 16, 2003.

\bibitem[Kemp and Tenenbaum(2008)]{kemp2008discovery}
Charles Kemp and Joshua~B Tenenbaum.
\newblock The discovery of structural form.
\newblock \emph{Proceedings of the National Academy of Sciences}, 105\penalty0 (31):\penalty0 10687--10692, 2008.

\bibitem[Lake et~al.(2015)Lake, Salakhutdinov, and Tenenbaum]{lake2015human}
Brenden~M Lake, Ruslan Salakhutdinov, and Joshua~B Tenenbaum.
\newblock Human-level concept learning through probabilistic program induction.
\newblock \emph{Science}, 350\penalty0 (6266):\penalty0 1332--1338, 2015.

\bibitem[Rule et~al.(2020)Rule, Tenenbaum, and Piantadosi]{rule2020child}
Joshua~S Rule, Joshua~B Tenenbaum, and Steven~T Piantadosi.
\newblock The child as hacker.
\newblock \emph{Trends in cognitive sciences}, 24\penalty0 (11):\penalty0 900--915, 2020.

\bibitem[Ellis et~al.(2021)Ellis, Wong, Nye, Sabl{\'e}-Meyer, and Morales]{ellis2021dreamcoder}
Kevin Ellis, Catherine Wong, Maxwell Nye, Mathias Sabl{\'e}-Meyer, and Lucas et~al Morales.
\newblock Dreamcoder: Bootstrapping inductive program synthesis with wake-sleep library learning.
\newblock In \emph{Proceedings of the 42nd acm sigplan international conference on programming language design and implementation}, pages 835--850, 2021.

\bibitem[Ullman and Tenenbaum(2020{\natexlab{a}})]{ullman2020bayesian}
Tomer~D Ullman and Joshua~B Tenenbaum.
\newblock Bayesian models of conceptual development: Learning as building models of the world.
\newblock \emph{Annual Review of Developmental Psychology}, 2\penalty0 (1):\penalty0 533--558, 2020{\natexlab{a}}.

\bibitem[Lieder et~al.(2019)Lieder, Chen, Krueger, and Griffiths]{lieder2019cognitive}
Falk Lieder, Owen~X Chen, Paul~M Krueger, and Thomas~L Griffiths.
\newblock Cognitive prostheses for goal achievement.
\newblock \emph{Nature human behaviour}, 3\penalty0 (10):\penalty0 1096--1106, 2019.

\bibitem[Lieder and Griffiths(2020)]{lieder2020resource}
Falk Lieder and Thomas~L Griffiths.
\newblock Resource-rational analysis: Understanding human cognition as the optimal use of limited computational resources.
\newblock \emph{Behavioral and brain sciences}, 43:\penalty0 e1, 2020.

\bibitem[Icard(2023)]{icard2023resource}
Thomas Icard.
\newblock Resource rationality.
\newblock Book manuscript, 2023.

\bibitem[Newell and Simon(1972)]{newell1972human}
Allen Newell and Herbert~A. Simon.
\newblock \emph{Human problem solving}.
\newblock Prentice-Hall, 1972.

\bibitem[Mattar and Lengyel(2022)]{mattar2022planning}
Marcelo~G Mattar and M{\'a}t{\'e} Lengyel.
\newblock Planning in the brain.
\newblock \emph{Neuron}, 110\penalty0 (6):\penalty0 914--934, 2022.

\bibitem[Ho et~al.(2022{\natexlab{a}})Ho, Abel, Correa, Littman, and Cohen]{ho2022people}
Mark~K. Ho, David Abel, Carlos~G. Correa, Michael~L. Littman, and Jonathan D. et~al Cohen.
\newblock People construct simplified mental representations to plan.
\newblock \emph{Nature}, 606\penalty0 (7912):\penalty0 129--136, 2022{\natexlab{a}}.

\bibitem[Jara-Ettinger et~al.(2016)Jara-Ettinger, Gweon, Schulz, and Tenenbaum]{jara2016naive}
Julian Jara-Ettinger, Hyowon Gweon, Laura~E Schulz, and Joshua~B Tenenbaum.
\newblock The na{\"\i}ve utility calculus: Computational principles underlying commonsense psychology.
\newblock \emph{Trends in cognitive sciences}, 20\penalty0 (8):\penalty0 589--604, 2016.

\bibitem[Baker et~al.(2011)Baker, Saxe, and Tenenbaum]{baker2011bayesian}
Chris Baker, Rebecca Saxe, and Joshua Tenenbaum.
\newblock Bayesian theory of mind: Modeling joint belief-desire attribution.
\newblock In \emph{Proceedings of the annual meeting of the cognitive science society}, volume~33, 2011.

\bibitem[Ho et~al.(2022{\natexlab{b}})Ho, Saxe, and Cushman]{ho2022planning}
Mark~K Ho, Rebecca Saxe, and Fiery Cushman.
\newblock Planning with theory of mind.
\newblock \emph{Trends in Cognitive Sciences}, 26\penalty0 (11):\penalty0 959--971, 2022{\natexlab{b}}.

\bibitem[Frank and Goodman(2012)]{frank2012predicting}
Michael~C Frank and Noah~D Goodman.
\newblock Predicting pragmatic reasoning in language games.
\newblock \emph{Science}, 336\penalty0 (6084):\penalty0 998--998, 2012.

\bibitem[Degen(2023)]{degen2023rational}
Judith Degen.
\newblock The rational speech act framework.
\newblock \emph{Annual Review of Linguistics}, 9:\penalty0 519--540, 2023.

\bibitem[Binz et~al.(2023)Binz, Dasgupta, Jagadish, Botvinick, and Wang]{binz2023meta}
Marcel Binz, Ishita Dasgupta, Akshay~K Jagadish, Matthew Botvinick, and Jane X et~al Wang.
\newblock Meta-learned models of cognition.
\newblock \emph{Behavioral and Brain Sciences}, pages 1--38, 2023.

\bibitem[Grant et~al.(2018)Grant, Finn, Levine, Darrell, and Griffiths]{grant2018recasting}
Erin Grant, Chelsea Finn, Sergey Levine, Trevor Darrell, and Thomas Griffiths.
\newblock Recasting gradient-based meta-learning as hierarchical bayes.
\newblock In \emph{6th International Conference on Learning Representations, ICLR 2018}, 2018.

\bibitem[Lake and Baroni(2023)]{lake2023human}
Brenden~M Lake and Marco Baroni.
\newblock Human-like systematic generalization through a meta-learning neural network.
\newblock \emph{Nature}, 623\penalty0 (7985):\penalty0 115--121, 2023.

\bibitem[Ho and Griffiths(2022)]{ho2022cognitive}
Mark~K Ho and Thomas~L Griffiths.
\newblock Cognitive science as a source of forward and inverse models of human decisions for robotics and control.
\newblock \emph{Annual Review of Control, Robotics, and Autonomous Systems}, 5:\penalty0 33--53, 2022.

\bibitem[Yang et~al.(2023)Yang, Folke, and Shafto]{yang2023inner}
Scott Cheng-Hsin Yang, Tomas Folke, and Patrick Shafto.
\newblock The inner loop of collective human--machine intelligence.
\newblock \emph{Topics in cognitive science}, 2023.

\bibitem[Steyvers and Kumar(2023)]{steyvers2023threePublished}
Mark Steyvers and Aakriti Kumar.
\newblock Three challenges for ai-assisted decision-making.
\newblock \emph{Perspectives on Psychological Science}, page 17456916231181102, 2023.

\bibitem[Chater and Manning(2006)]{chater2006probabilistic}
Nick Chater and Christopher~D Manning.
\newblock Probabilistic models of language processing and acquisition.
\newblock \emph{Trends in cognitive sciences}, 10\penalty0 (7):\penalty0 335--344, 2006.

\bibitem[Oaksford and Chater(2007)]{oaksford2007bayesian}
Mike Oaksford and Nick Chater.
\newblock \emph{Bayesian rationality: The probabilistic approach to human reasoning}.
\newblock Oxford University Press, 2007.

\bibitem[Xu and Tenenbaum(2007)]{xu2007word}
Fei Xu and Joshua~B Tenenbaum.
\newblock Word learning as bayesian inference.
\newblock \emph{Psychological review}, 114\penalty0 (2):\penalty0 245, 2007.

\bibitem[Kersten et~al.(2004)Kersten, Mamassian, and Yuille]{kersten2004object}
Daniel Kersten, Pascal Mamassian, and Alan Yuille.
\newblock Object perception as bayesian inference.
\newblock \emph{Annu. Rev. Psychol.}, 55:\penalty0 271--304, 2004.

\bibitem[Yildirim et~al.(2020)Yildirim, Belledonne, Freiwald, and Tenenbaum]{yildirim2020efficient}
Ilker Yildirim, Mario Belledonne, Winrich Freiwald, and Josh Tenenbaum.
\newblock Efficient inverse graphics in biological face processing.
\newblock \emph{Science advances}, 6\penalty0 (10):\penalty0 eaax5979, 2020.

\bibitem[Allen et~al.(2020)Allen, Smith, and Tenenbaum]{allen2020rapid}
Kelsey~R. Allen, Kevin~A. Smith, and Joshua~B. Tenenbaum.
\newblock Rapid trial-and-error learning with simulation supports flexible tool use and physical reasoning.
\newblock \emph{Proceedings of the National Academy of Sciences}, 117\penalty0 (47):\penalty0 29302--29310, 2020.

\bibitem[Zhang et~al.(2023{\natexlab{b}})Zhang, Wong, Grand, and Tenenbaum]{zhang2023grounded}
Cedegao~E Zhang, Lionel Wong, Gabriel Grand, and Joshua~B Tenenbaum.
\newblock Grounded physical language understanding with probabilistic programs and simulated worlds.
\newblock In \emph{Proceedings of the Annual Meeting of the Cognitive Science Society}, volume~45, 2023{\natexlab{b}}.

\bibitem[Tenenbaum(1998)]{tenenbaum1998bayesian}
Joshua Tenenbaum.
\newblock Bayesian modeling of human concept learning.
\newblock \emph{Advances in neural information processing systems}, 11, 1998.

\bibitem[Goodman et~al.(2008{\natexlab{b}})Goodman, Tenenbaum, Feldman, and Griffiths]{goodman2008rational}
Noah~D Goodman, Joshua~B Tenenbaum, Jacob Feldman, and Thomas~L Griffiths.
\newblock A rational analysis of rule-based concept learning.
\newblock \emph{Cognitive science}, 32\penalty0 (1):\penalty0 108--154, 2008{\natexlab{b}}.

\bibitem[Piantadosi et~al.(2016)Piantadosi, Tenenbaum, and Goodman]{piantadosi2016logical}
Steven~T Piantadosi, Joshua~B Tenenbaum, and Noah~D Goodman.
\newblock The logical primitives of thought: Empirical foundations for compositional cognitive models.
\newblock \emph{Psychological review}, 123\penalty0 (4):\penalty0 392, 2016.

\bibitem[Griffiths et~al.(2004)Griffiths, Steyvers, Blei, and Tenenbaum]{griffiths2004integrating}
Thomas Griffiths, Mark Steyvers, David Blei, and Joshua Tenenbaum.
\newblock Integrating topics and syntax.
\newblock \emph{Advances in neural information processing systems}, 17, 2004.

\bibitem[Goodman and Lassiter(2015)]{goodman2015probabilistic}
Noah~D Goodman and Daniel Lassiter.
\newblock Probabilistic semantics and pragmatics uncertainty in language and thought.
\newblock \emph{The handbook of contemporary semantic theory}, pages 655--686, 2015.

\bibitem[Yang and Piantadosi(2022)]{yang2022one}
Yuan Yang and Steven~T Piantadosi.
\newblock One model for the learning of language.
\newblock \emph{Proceedings of the National Academy of Sciences}, 119\penalty0 (5):\penalty0 e2021865119, 2022.

\bibitem[Schulz et~al.(2007)Schulz, Bonawitz, and Griffiths]{schulz2007can}
Laura~E Schulz, Elizabeth~Baraff Bonawitz, and Thomas~L Griffiths.
\newblock Can being scared cause tummy aches? naive theories, ambiguous evidence, and preschoolers' causal inferences.
\newblock \emph{Developmental psychology}, 43\penalty0 (5):\penalty0 1124, 2007.

\bibitem[Gopnik et~al.(2004)Gopnik, Glymour, Sobel, Schulz, and Kushnir]{gopnik2004theory}
Alison Gopnik, Clark Glymour, David~M Sobel, Laura~E Schulz, and Tamar et~al Kushnir.
\newblock A theory of causal learning in children: causal maps and bayes nets.
\newblock \emph{Psychological review}, 111\penalty0 (1):\penalty0 3, 2004.

\bibitem[Kirfel et~al.(2022)Kirfel, Icard, and Gerstenberg]{kirfel2022inference}
Lara Kirfel, Thomas Icard, and Tobias Gerstenberg.
\newblock Inference from explanation.
\newblock \emph{Journal of Experimental Psychology: General}, 151\penalty0 (7):\penalty0 1481, 2022.

\bibitem[Lagnado et~al.(2013)Lagnado, Gerstenberg, and Zultan]{lagnado2013causal}
David~A Lagnado, Tobias Gerstenberg, and Ro'i Zultan.
\newblock Causal responsibility and counterfactuals.
\newblock \emph{Cognitive science}, 37\penalty0 (6):\penalty0 1036--1073, 2013.

\bibitem[Hemmer and Steyvers(2009)]{hemmer2009bayesian}
Pernille Hemmer and Mark Steyvers.
\newblock A bayesian account of reconstructive memory.
\newblock \emph{Topics in Cognitive Science}, 1\penalty0 (1):\penalty0 189--202, 2009.

\bibitem[Ullman and Tenenbaum(2020{\natexlab{b}})]{ullman2020}
Tomer~D. Ullman and Joshua~B. Tenenbaum.
\newblock Bayesian models of conceptual development: Learning as building models of the world.
\newblock \emph{Annual Review of Developmental Psychology}, 2\penalty0 (1):\penalty0 533--558, 2020{\natexlab{b}}.
\newblock \doi{10.1146/annurev-devpsych-121318-084833}.

\bibitem[Griffiths and Tenenbaum(2009)]{griffiths2009theory}
Thomas~L Griffiths and Joshua~B Tenenbaum.
\newblock Theory-based causal induction.
\newblock \emph{Psychological review}, 116\penalty0 (4):\penalty0 661, 2009.

\bibitem[Vul et~al.(2014)Vul, Goodman, Griffiths, and Tenenbaum]{vul2014one}
Edward Vul, Noah Goodman, Thomas~L Griffiths, and Joshua~B Tenenbaum.
\newblock One and done? optimal decisions from very few samples.
\newblock \emph{Cognitive science}, 38\penalty0 (4):\penalty0 599--637, 2014.

\bibitem[Hay et~al.(2014)Hay, Russell, Tolpin, and Shimony]{hay2014selecting}
Nicholas Hay, Stuart Russell, David Tolpin, and Solomon~Eyal Shimony.
\newblock Selecting computations: Theory and applications.
\newblock \emph{arXiv preprint arXiv:1408.2048}, 2014.

\bibitem[Tomov et~al.(2020)Tomov, Yagati, Kumar, Yang, and Gershman]{tomov2020}
Momchil~S. Tomov, Samyukta Yagati, Agni Kumar, Wanqian Yang, and Samuel~J. Gershman.
\newblock Discovery of hierarchical representations for efficient planning.
\newblock \emph{PLOS Computational Biology}, 16\penalty0 (4):\penalty0 1--42, 04 2020.
\newblock \doi{10.1371/journal.pcbi.1007594}.

\bibitem[Baker and Tenenbaum(2014)]{baker2014modeling}
Chris~L Baker and Joshua~B Tenenbaum.
\newblock Modeling human plan recognition using bayesian theory of mind.
\newblock \emph{Plan, activity, and intent recognition: Theory and practice}, 7:\penalty0 177--204, 2014.

\bibitem[Callaway et~al.(2022)Callaway, van Opheusden, Gul, Das, and Krueger]{callaway2022rational}
Frederick Callaway, Bas van Opheusden, Sayan Gul, Priyam Das, and Paul M et~al Krueger.
\newblock Rational use of cognitive resources in human planning.
\newblock \emph{Nature Human Behaviour}, 6\penalty0 (8):\penalty0 1112--1125, 2022.

\bibitem[Baker et~al.(2017)Baker, Jara-Ettinger, Saxe, and Tenenbaum]{baker2017rational}
Chris~L Baker, Julian Jara-Ettinger, Rebecca Saxe, and Joshua~B Tenenbaum.
\newblock Rational quantitative attribution of beliefs, desires and percepts in human mentalizing.
\newblock \emph{Nature Human Behaviour}, 1\penalty0 (4):\penalty0 0064, 2017.

\bibitem[Zhi-Xuan et~al.(2020)Zhi-Xuan, Mann, Silver, Tenenbaum, and Mansinghka]{zhi2020online}
Tan Zhi-Xuan, Jordyn Mann, Tom Silver, Josh Tenenbaum, and Vikash Mansinghka.
\newblock Online bayesian goal inference for boundedly rational planning agents.
\newblock \emph{Advances in neural information processing systems}, 33:\penalty0 19238--19250, 2020.

\bibitem[Ying et~al.(2023)Ying, Collins, Wei, Zhang, and Zhi-Xuan]{ying2023neuro}
Lance Ying, Katherine~M Collins, Megan Wei, Cedegao~E Zhang, and Tan et~al Zhi-Xuan.
\newblock The neuro-symbolic inverse planning engine (nipe): Modeling probabilistic social inferences from linguistic inputs.
\newblock \emph{arXiv preprint arXiv:2306.14325}, 2023.

\bibitem[Johnson-Laird(1983)]{johnson1983mental}
Philip~Nicholas Johnson-Laird.
\newblock \emph{Mental models: Towards a cognitive science of language, inference, and consciousness}.
\newblock Number~6. Harvard University Press, 1983.

\bibitem[Byrne(2002)]{byrne2002mental}
Ruth~MJ Byrne.
\newblock Mental models and counterfactual thoughts about what might have been.
\newblock \emph{Trends in cognitive sciences}, 6\penalty0 (10):\penalty0 426--431, 2002.

\bibitem[Shafto et~al.(2014)Shafto, Goodman, and Griffiths]{shafto2014rational}
Patrick Shafto, Noah~D Goodman, and Thomas~L Griffiths.
\newblock A rational account of pedagogical reasoning: Teaching by, and learning from, examples.
\newblock \emph{Cognitive psychology}, 71:\penalty0 55--89, 2014.

\bibitem[Sumers et~al.(2021)Sumers, Ho, Hawkins, Narasimhan, and Griffiths]{sumers2021learning}
Theodore~R Sumers, Mark~K Ho, Robert~D Hawkins, Karthik Narasimhan, and Thomas~L Griffiths.
\newblock Learning rewards from linguistic feedback.
\newblock In \emph{Proceedings of the AAAI Conference on Artificial Intelligence}, volume~35, pages 6002--6010, 2021.

\bibitem[Liquin et~al.(2023)Liquin, Luzuriaga, and Gureckis]{liquin2023teaching}
Emily~G Liquin, Nicole Luzuriaga, and Todd~M Gureckis.
\newblock Teaching and learning through pedagogical environment design.
\newblock In \emph{Proceedings of the Annual Meeting of the Cognitive Science Society}, volume~45, 2023.

\bibitem[Kumar et~al.(2023)Kumar, Smyth, and Steyvers]{kumar2023differentiating}
Aakriti Kumar, Padhraic Smyth, and Mark Steyvers.
\newblock Differentiating mental models of self et al: A hierarchical framework for knowledge assessment.
\newblock \emph{Psychological Review}, 2023.

\bibitem[Hawkins et~al.(2023{\natexlab{a}})Hawkins, Franke, Frank, Goldberg, and Smith]{hawkins2023partners}
Robert~D Hawkins, Michael Franke, Michael~C Frank, Adele~E Goldberg, and Kenny et~al Smith.
\newblock From partners to populations: A hierarchical bayesian account of coordination and convention.
\newblock \emph{Psychological Review}, 130\penalty0 (4):\penalty0 977, 2023{\natexlab{a}}.

\bibitem[Hawkins et~al.(2023{\natexlab{b}})Hawkins, Berdahl, Pentland, Tenenbaum, and Goodman]{hawkins2023flexible}
Robert~D Hawkins, Andrew~M Berdahl, Alex~‘Sandy’ Pentland, Joshua~B Tenenbaum, and Noah D et~al Goodman.
\newblock Flexible social inference facilitates targeted social learning when rewards are not observable.
\newblock \emph{Nature Human Behaviour}, pages 1--10, 2023{\natexlab{b}}.

\bibitem[Goodman and Stuhlm{\"u}ller(2013)]{goodman2013knowledge}
Noah~D Goodman and Andreas Stuhlm{\"u}ller.
\newblock Knowledge and implicature: Modeling language understanding as social cognition.
\newblock \emph{Topics in cognitive science}, 5\penalty0 (1):\penalty0 173--184, 2013.

\bibitem[Ho et~al.(2021)Ho, Cushman, Littman, and Austerweil]{ho2021communication}
Mark~K Ho, Fiery Cushman, Michael~L Littman, and Joseph~L Austerweil.
\newblock Communication in action: Planning and interpreting communicative demonstrations.
\newblock \emph{Journal of Experimental Psychology: General}, 150\penalty0 (11):\penalty0 2246, 2021.

\bibitem[Griffiths(2020)]{griffiths2020understanding}
Thomas~L Griffiths.
\newblock Understanding human intelligence through human limitations.
\newblock \emph{Trends in Cognitive Sciences}, 24\penalty0 (11):\penalty0 873--883, 2020.

\bibitem[Tversky and Kahneman(1973)]{tversky1973availability}
Amos Tversky and Daniel Kahneman.
\newblock Availability: A heuristic for judging frequency and probability.
\newblock \emph{Cognitive psychology}, 5\penalty0 (2):\penalty0 207--232, 1973.

\bibitem[Tversky and Kahneman(1974)]{tversky1974judgment}
Amos Tversky and Daniel Kahneman.
\newblock Judgment under uncertainty: Heuristics and biases: Biases in judgments reveal some heuristics of thinking under uncertainty.
\newblock \emph{science}, 185\penalty0 (4157):\penalty0 1124--1131, 1974.

\bibitem[Zhu et~al.(2023)Zhu, Sundh, Spicer, Chater, and Sanborn]{zhu2023autocorrelated}
Jian-Qiao Zhu, Joakim Sundh, Jake Spicer, Nick Chater, and Adam~N Sanborn.
\newblock The autocorrelated bayesian sampler: A rational process for probability judgments, estimates, confidence intervals, choices, confidence judgments, and response times.
\newblock \emph{Psychological Review}, 2023.

\bibitem[Van~Rooij(2008)]{van2008tractable}
Iris Van~Rooij.
\newblock The tractable cognition thesis.
\newblock \emph{Cognitive science}, 32\penalty0 (6):\penalty0 939--984, 2008.

\bibitem[Icard and Goodman(2015)]{icard2015resource}
Thomas Icard and Noah~D Goodman.
\newblock A resource-rational approach to the causal frame problem.
\newblock In \emph{CogSci}, 2015.

\bibitem[Anderson(1990)]{anderson1990adaptive}
John~R Anderson.
\newblock \emph{The adaptive character of thought}.
\newblock Psychology Press, 1990.

\bibitem[Cheyette et~al.(2023)Cheyette, Callaway, Bramley, Nelson, and Tenenbaum]{cheyette2023people}
Samuel~J Cheyette, Frederick Callaway, Neil~R Bramley, Jonathan~D Nelson, and Josh Tenenbaum.
\newblock People seek easily interpretable information.
\newblock In \emph{Proceedings of the Annual Meeting of the Cognitive Science Society}, volume~45, 2023.

\bibitem[Saad(2022)]{saad2022scalable}
Feras Ahmad~Khaled Saad.
\newblock \emph{Scalable Structure Learning, Inference, and Analysis with Probabilistic Programs}.
\newblock PhD thesis, Massachusetts Institute of Technology, 2022.

\bibitem[Lew et~al.(2020)Lew, Tessler, Mansinghka, and Tenenbaum]{lew2020leveraging}
Alexander~K Lew, Michael~Henry Tessler, Vikash~K Mansinghka, and Joshua~B Tenenbaum.
\newblock Leveraging unstructured statistical knowledge in a probabilistic language of thought.
\newblock In \emph{Proceedings of the annual conference of the cognitive science society}, 2020.

\bibitem[Gothoskar et~al.(2023)Gothoskar, Ghavami, Li, Curtis, and Noseworthy]{gothoskar2023bayes3d}
Nishad Gothoskar, Matin Ghavami, Eric Li, Aidan Curtis, and Michael et~al Noseworthy.
\newblock Bayes3d: fast learning and inference in structured generative models of 3d objects and scenes.
\newblock \emph{arXiv preprint arXiv:2312.08715}, 2023.

\bibitem[Mansinghka et~al.(2018)Mansinghka, Schaechtle, Handa, Radul, Chen, and Rinard]{mansinghka2018probabilistic}
Vikash~K Mansinghka, Ulrich Schaechtle, Shivam Handa, Alexey Radul, Yutian Chen, and Martin Rinard.
\newblock Probabilistic programming with programmable inference.
\newblock In \emph{Proceedings of the 39th ACM SIGPLAN Conference on Programming Language Design and Implementation}, pages 603--616, 2018.

\bibitem[Lew et~al.(2023{\natexlab{a}})Lew, Huot, Staton, and Mansinghka]{adevLew2023}
Alexander~K. Lew, Mathieu Huot, Sam Staton, and Vikash~K. Mansinghka.
\newblock Adev: Sound automatic differentiation of expected values of probabilistic programs.
\newblock \emph{Proc. ACM Program. Lang.}, 7\penalty0 (POPL), jan 2023{\natexlab{a}}.
\newblock \doi{10.1145/3571198}.
\newblock URL \url{https://doi.org/10.1145/3571198}.

\bibitem[Becker et~al.(2024)Becker, Lew, Wang, Ghavami, Huot, Rinard, and Mansinghka]{becker2024probabilistic}
McCoy~R Becker, Alexander~K Lew, Xiaoyan Wang, Matin Ghavami, Mathieu Huot, Martin~C Rinard, and Vikash~K Mansinghka.
\newblock Probabilistic programming with programmable variational inference.
\newblock \emph{Proceedings of the ACM on Programming Languages}, 8\penalty0 (PLDI):\penalty0 2123--2147, 2024.

\bibitem[Saad et~al.(2021)Saad, Rinard, and Mansinghka]{saad2021sppl}
Feras~A Saad, Martin~C Rinard, and Vikash~K Mansinghka.
\newblock Sppl: probabilistic programming with fast exact symbolic inference.
\newblock In \emph{Proceedings of the 42nd acm sigplan international conference on programming language design and implementation}, pages 804--819, 2021.

\bibitem[Lew et~al.(2023{\natexlab{b}})Lew, Ghavamizadeh, Rinard, and Mansinghka]{lew2023probabilistic}
Alexander~K Lew, Matin Ghavamizadeh, Martin~C Rinard, and Vikash~K Mansinghka.
\newblock Probabilistic programming with stochastic probabilities.
\newblock \emph{Proceedings of the ACM on Programming Languages}, 7\penalty0 (PLDI):\penalty0 1708--1732, 2023{\natexlab{b}}.

\bibitem[Guggenberger et~al.(2020)Guggenberger, M{\"o}ller, Haarhaus, G{\"u}r, and Otto]{guggenberger2020ecosystem}
Tobias~Moritz Guggenberger, Frederik M{\"o}ller, Tim Haarhaus, Inan G{\"u}r, and Boris Otto.
\newblock Ecosystem types in information systems.
\newblock In \emph{Twenty-Eighth European Conference on Information Systems (ECIS2020)}, 2020.

\bibitem[Goodman and Flaxman(2017)]{goodman2017european}
Bryce Goodman and Seth Flaxman.
\newblock European union regulations on algorithmic decision-making and a “right to explanation”.
\newblock \emph{AI magazine}, 38\penalty0 (3):\penalty0 50--57, 2017.

\bibitem[Wachter and Mittelstadt(2019)]{wachter2019right}
Sandra Wachter and Brent Mittelstadt.
\newblock A right to reasonable inferences: re-thinking data protection law in the age of big data and ai.
\newblock \emph{Colum. Bus. L. Rev.}, page 494, 2019.

\bibitem[Fui-Hoon~Nah et~al.(2023)Fui-Hoon~Nah, Zheng, Cai, Siau, and Chen]{fui2023generative}
Fiona Fui-Hoon~Nah, Ruilin Zheng, Jingyuan Cai, Keng Siau, and Langtao Chen.
\newblock Generative ai and chatgpt: Applications, challenges, and ai-human collaboration, 2023.

\bibitem[Norman(1988)]{norman1988design}
Don Norman.
\newblock \emph{Design Of Everyday Things}.
\newblock New York: Basic Books. Olins, W.(2005). A Marca. Lisboa: Verbo. Packard, V~…, 1988.

\bibitem[Chemero(2018)]{chemero2018outline}
Anthony Chemero.
\newblock An outline of a theory of affordances.
\newblock In \emph{How Shall Affordances Be Refined?}, pages 181--195. Routledge, 2018.

\bibitem[Zerilli et~al.(2022{\natexlab{a}})Zerilli, Bhatt, and Weller]{zerilli2022transparency}
John Zerilli, Umang Bhatt, and Adrian Weller.
\newblock How transparency modulates trust in artificial intelligence.
\newblock \emph{Patterns}, page 100455, 2022{\natexlab{a}}.

\bibitem[Messeri and Crockett(2024)]{messeri2024artificial}
Lisa Messeri and MJ~Crockett.
\newblock Artificial intelligence and illusions of understanding in scientific research.
\newblock \emph{Nature}, 627\penalty0 (8002):\penalty0 49--58, 2024.

\bibitem[Tejeda et~al.(2022)Tejeda, Kumar, Smyth, and Steyvers]{tejeda2022ai}
Heliodoro Tejeda, Aakriti Kumar, Padhraic Smyth, and Mark Steyvers.
\newblock Ai-assisted decision-making: A cognitive modeling approach to infer latent reliance strategies.
\newblock \emph{Computational Brain \& Behavior}, 5\penalty0 (4):\penalty0 1--18, 2022.

\bibitem[Steyvers et~al.(2022)Steyvers, Tejeda, Kerrigan, and Smyth]{steyvers2022bayesian}
Mark Steyvers, Heliodoro Tejeda, Gavin Kerrigan, and Padhraic Smyth.
\newblock Bayesian modeling of human--ai complementarity.
\newblock \emph{Proceedings of the National Academy of Sciences}, 119\penalty0 (11):\penalty0 e2111547119, 2022.

\bibitem[Chandra et~al.(2024{\natexlab{c}})Chandra, Chen, Li, Ragan-Kelley, and Tenenbaum]{chandra2024cooperative}
Kartik Chandra, Tony Chen, Tzu-Mao Li, Jonathan Ragan-Kelley, and Josh Tenenbaum.
\newblock Cooperative explanation as rational communication.
\newblock In \emph{Proceedings of the Annual Meeting of the Cognitive Science Society}, volume~46, 2024{\natexlab{c}}.

\bibitem[Hadfield-Menell et~al.(2016)Hadfield-Menell, Russell, Abbeel, and Dragan]{hadfield2016cooperative}
Dylan Hadfield-Menell, Stuart~J Russell, Pieter Abbeel, and Anca Dragan.
\newblock Cooperative inverse reinforcement learning.
\newblock \emph{Advances in neural information processing systems}, 29, 2016.

\bibitem[Chandra et~al.(2023)Chandra, Li, Tenenbaum, and Ragan-Kelley]{chandra2023acting}
Kartik Chandra, Tzu-Mao Li, Joshua Tenenbaum, and Jonathan Ragan-Kelley.
\newblock Acting as inverse inverse planning.
\newblock In \emph{ACM SIGGRAPH 2023 Conference Proceedings}, pages 1--12, 2023.

\bibitem[Chen et~al.(2024)Chen, Houlihan, Chandra, Tenenbaum, and Saxe]{chen2024intervening}
Tony Chen, Sean~Dae Houlihan, Kartik Chandra, Josh Tenenbaum, and Rebecca Saxe.
\newblock Intervening on emotions by planning over a theory of mind.
\newblock In \emph{Proceedings of the Annual Meeting of the Cognitive Science Society}, volume~46, 2024.

\bibitem[Chandra et~al.(2024{\natexlab{d}})Chandra, Li, Tenenbaum, and Ragan-Kelley]{chandra2024storytelling}
Kartik Chandra, Tzu-Mao Li, Joshua~B Tenenbaum, and Jonathan Ragan-Kelley.
\newblock Storytelling as inverse inverse planning.
\newblock \emph{Topics in Cognitive Science}, 16\penalty0 (1):\penalty0 54--70, 2024{\natexlab{d}}.

\bibitem[Loula et~al.(2024)Loula, Collins, Schaechtle, Tenenbaum, and Weller]{loula2024learning}
Jo{\~a}o Loula, Katherine~M Collins, Ulrich Schaechtle, Joshua~B Tenenbaum, and Adrian et~al Weller.
\newblock Learning generative population models from multiple clinical datasets via probabilistic programming.
\newblock In \emph{ICML 2024 Workshop on Efficient and Accessible Foundation Models for Biological Discovery}, 2024.

\bibitem[Cabitza et~al.(2017)Cabitza, Rasoini, and Gensini]{cabitza2017unintended}
Federico Cabitza, Raffaele Rasoini, and Gian~Franco Gensini.
\newblock Unintended consequences of machine learning in medicine.
\newblock \emph{Jama}, 318\penalty0 (6):\penalty0 517--518, 2017.

\bibitem[Mozannar and Sontag(2020)]{mozannar2020consistent}
Hussein Mozannar and David Sontag.
\newblock Consistent estimators for learning to defer to an expert.
\newblock In \emph{International Conference on Machine Learning}, pages 7076--7087. PMLR, 2020.

\bibitem[Dvijotham et~al.(2023)Dvijotham, Winkens, Barsbey, Ghaisas, and Stanforth]{dvijotham2023enhancing}
Krishnamurthy Dvijotham, Jim Winkens, Melih Barsbey, Sumedh Ghaisas, and Robert et~al Stanforth.
\newblock Enhancing the reliability and accuracy of ai-enabled diagnosis via complementarity-driven deferral to clinicians.
\newblock \emph{Nature Medicine}, pages 1--7, 2023.

\bibitem[Tsvetkova et~al.(2024)Tsvetkova, Yasseri, Pescetelli, and Werner]{tsvetkova2024human}
Milena Tsvetkova, Taha Yasseri, Niccolo Pescetelli, and Tobias Werner.
\newblock Human-machine social systems.
\newblock \emph{arXiv preprint arXiv:2402.14410}, 2024.

\bibitem[Schneiders et~al.(2022)Schneiders, Cheon, Kjeldskov, Rehm, and Skov]{schneiders2022non}
Eike Schneiders, EunJeong Cheon, Jesper Kjeldskov, Matthias Rehm, and Mikael~B Skov.
\newblock Non-dyadic interaction: A literature review of 15 years of human-robot interaction conference publications.
\newblock \emph{ACM Transactions on Human-Robot Interaction (THRI)}, 11\penalty0 (2):\penalty0 1--32, 2022.

\bibitem[Hornecker et~al.(2022)Hornecker, Krummheuer, Bischof, and Rehm]{hornecker2022beyond}
Eva Hornecker, Antonia Krummheuer, Andreas Bischof, and Matthias Rehm.
\newblock Beyond dyadic hri: Building robots for society.
\newblock \emph{interactions}, 29\penalty0 (3):\penalty0 48--53, 2022.

\bibitem[Yadav and Mehta(2024)]{yadav2024beyond}
Aakash Yadav and Ranjana Mehta.
\newblock Beyond dyadic interactions: Assessing trust networks in multi-human-robot teams.
\newblock In \emph{Companion of the 2024 ACM/IEEE International Conference on Human-Robot Interaction}, pages 1153--1157, 2024.

\bibitem[Sucholutsky et~al.(2024)Sucholutsky, Collins, Malaviya, Jacoby, and Liu]{sucholutsky2024representationalTeaching}
Ilia Sucholutsky, Katherine~M Collins, Maya Malaviya, Nori Jacoby, and Weiyang et~al Liu.
\newblock Representational alignment supports effective machine teaching.
\newblock \emph{arXiv Preprint arXiv:2406.04302}, 2024.

\bibitem[Li et~al.(2023)Li, Li, Ouyang, Mo, and Ren]{li2023three}
Ling Li, Xiaojian Li, Bo~Ouyang, Hangjie Mo, and Hongliang et~al Ren.
\newblock Three-dimensional collision avoidance method for robot-assisted minimally invasive surgery.
\newblock \emph{Cyborg and Bionic Systems}, 4:\penalty0 0042, 2023.

\bibitem[Boyce et~al.(2024)Boyce, Hawkins, Goodman, and Frank]{boyce2024interaction}
Veronica Boyce, Robert~D Hawkins, Noah~D Goodman, and Michael~C Frank.
\newblock Interaction structure constrains the emergence of conventions in group communication.
\newblock \emph{Proceedings of the National Academy of Sciences}, 121\penalty0 (28), 2024.

\bibitem[Trouille et~al.(2019)Trouille, Lintott, and Fortson]{trouille2019citizen}
Laura Trouille, Chris~J Lintott, and Lucy~F Fortson.
\newblock Citizen science frontiers: Efficiency, engagement, and serendipitous discovery with human--machine systems.
\newblock \emph{Proceedings of the National Academy of Sciences}, 116\penalty0 (6):\penalty0 1902--1909, 2019.

\bibitem[Hornb{\ae}k and Oulasvirta(2017)]{hornbaek2017interaction}
Kasper Hornb{\ae}k and Antti Oulasvirta.
\newblock What is interaction?
\newblock In \emph{Proceedings of the 2017 CHI Conference on Human Factors in Computing Systems}, pages 5040--5052, 2017.

\bibitem[Lee et~al.(2023{\natexlab{b}})Lee, Srivastava, Hardy, Thickstun, and Durmus]{lee2023evaluating}
Mina Lee, Megha Srivastava, Amelia Hardy, John Thickstun, and Esin et~al Durmus.
\newblock Evaluating human-language model interaction.
\newblock \emph{Transactions on Machine Learning Research}, 2023{\natexlab{b}}.

\bibitem[Allen et~al.(2024)Allen, Br{\"a}ndle, Botvinick, Fan, and Gershman]{allen2024using}
Kelsey Allen, Franziska Br{\"a}ndle, Matthew Botvinick, Judith~E Fan, and Samuel J et~al Gershman.
\newblock Using games to understand the mind.
\newblock \emph{Nature Human Behaviour}, pages 1--9, 2024.

\bibitem[Park et~al.(2023)Park, O'Brien, Cai, Morris, and et~al]{park2023generative}
Joon~Sung Park, Joseph~C. O'Brien, Carrie~J. Cai, Meredith~Ringel Morris, and Percy~Liang et~al.
\newblock Generative agents: Interactive simulacra of human behavior, 2023.

\bibitem[Brown and Sandholm(2019)]{brown2019superhuman}
Noam Brown and Tuomas Sandholm.
\newblock Superhuman ai for multiplayer poker.
\newblock \emph{Science}, 365\penalty0 (6456):\penalty0 885--890, 2019.

\bibitem[Bakhtin et~al.(2022)Bakhtin, Brown, Dinan, Farina, and Flaherty]{meta2022human}
Anton Bakhtin, Noam Brown, Emily Dinan, Gabriele Farina, and Colin et~al Flaherty.
\newblock Human-level play in the game of diplomacy by combining language models with strategic reasoning.
\newblock \emph{Science}, 378\penalty0 (6624):\penalty0 1067--1074, 2022.

\bibitem[Logg et~al.(2019)Logg, Minson, and Moore]{logg2019algorithm}
Jennifer~M Logg, Julia~A Minson, and Don~A Moore.
\newblock Algorithm appreciation: People prefer algorithmic to human judgment.
\newblock \emph{Organizational Behavior and Human Decision Processes}, 151:\penalty0 90--103, 2019.

\bibitem[Green and Chen(2019)]{green2019principles}
Ben Green and Yiling Chen.
\newblock The principles and limits of algorithm-in-the-loop decision making.
\newblock \emph{Proceedings of the ACM on Human-Computer Interaction}, 3\penalty0 (CSCW):\penalty0 1--24, 2019.

\bibitem[Inuwa-Dutse et~al.(2023)Inuwa-Dutse, Toniolo, Weller, and Bhatt]{inuwa2023algorithmic}
Isa Inuwa-Dutse, Alice Toniolo, Adrian Weller, and Umang Bhatt.
\newblock Algorithmic loafing and mitigation strategies in human-ai teams.
\newblock \emph{Computers in Human Behavior: Artificial Humans}, 1\penalty0 (2):\penalty0 100024, 2023.

\bibitem[Hofman et~al.(2023)Hofman, Goldstein, and Rothschild]{hofman2023steroids}
Jake~M Hofman, Daniel~G Goldstein, and David~M Rothschild.
\newblock Steroids, sneakers, coach: The spectrum of human-ai relationships.
\newblock \emph{Available at SSRN 4578180}, 2023.

\bibitem[Buschek et~al.(2021)Buschek, Z\"{u}rn, and Eiband]{buschek2021impact}
Daniel Buschek, Martin Z\"{u}rn, and Malin Eiband.
\newblock The impact of multiple parallel phrase suggestions on email input and composition behaviour of native and non-native english writers.
\newblock In \emph{Proceedings of the 2021 CHI Conference on Human Factors in Computing Systems}, CHI '21, New York, NY, USA, 2021. Association for Computing Machinery.
\newblock ISBN 9781450380966.
\newblock \doi{10.1145/3411764.3445372}.

\bibitem[Bu{\c{c}}inca et~al.(2021)Bu{\c{c}}inca, Malaya, and Gajos]{buccinca2021trust}
Zana Bu{\c{c}}inca, Maja~Barbara Malaya, and Krzysztof~Z Gajos.
\newblock To trust or to think: cognitive forcing functions can reduce overreliance on ai in ai-assisted decision-making.
\newblock \emph{Proceedings of the ACM on Human-Computer Interaction}, 5\penalty0 (CSCW1):\penalty0 1--21, 2021.

\bibitem[Dietvorst et~al.(2015)Dietvorst, Simmons, and Massey]{dietvorst2015algorithm}
Berkeley~J Dietvorst, Joseph~P Simmons, and Cade Massey.
\newblock Algorithm aversion: people erroneously avoid algorithms after seeing them err.
\newblock \emph{Journal of Experimental Psychology: General}, 144\penalty0 (1):\penalty0 114, 2015.

\bibitem[Dietvorst et~al.(2018)Dietvorst, Simmons, and Massey]{dietvorst2018overcoming}
Berkeley~J Dietvorst, Joseph~P Simmons, and Cade Massey.
\newblock Overcoming algorithm aversion: People will use imperfect algorithms if they can (even slightly) modify them.
\newblock \emph{Management science}, 64\penalty0 (3):\penalty0 1155--1170, 2018.

\bibitem[Zerilli et~al.(2022{\natexlab{b}})Zerilli, Bhatt, and Weller]{zerilli2021how}
John Zerilli, Umang Bhatt, and Adrian Weller.
\newblock {T}ransparency {M}odulates {T}rust in {A}rtificial {I}ntelligence.
\newblock \emph{Patterns}, 2022{\natexlab{b}}.

\bibitem[Mumford(1936)]{mumford2010technics}
Lewis Mumford.
\newblock \emph{Technics and civilization}.
\newblock 1936.

\bibitem[Weizenbaum(1976)]{weizenbaum1976computer}
Joseph Weizenbaum.
\newblock Computer power and human reason: From judgment to calculation.
\newblock 1976.

\bibitem[Weidinger et~al.(2022)Weidinger, Uesato, Rauh, Griffin, and Huang]{weidinger2022taxonomy}
Laura Weidinger, Jonathan Uesato, Maribeth Rauh, Conor Griffin, and Po-Sen et~al Huang.
\newblock Taxonomy of risks posed by language models.
\newblock In \emph{Proceedings of the 2022 ACM Conference on Fairness, Accountability, and Transparency}, pages 214--229, 2022.

\bibitem[Shneiderman(2022{\natexlab{b}})]{shneiderman2022human}
Ben Shneiderman.
\newblock \emph{Human-centered AI}.
\newblock Oxford University Press, 2022{\natexlab{b}}.

\bibitem[Zhuang and Hadfield-Menell(2020)]{zhuang2020consequences}
Simon Zhuang and Dylan Hadfield-Menell.
\newblock Consequences of misaligned ai.
\newblock \emph{Advances in Neural Information Processing Systems}, 33:\penalty0 15763--15773, 2020.

\bibitem[Kalai and Vempala(2023)]{kalai2023calibrated}
Adam~Tauman Kalai and Santosh~S Vempala.
\newblock Calibrated language models must hallucinate.
\newblock \emph{arXiv preprint arXiv:2311.14648}, 2023.

\bibitem[Amodei et~al.(2016)Amodei, Olah, Steinhardt, Christiano, and Schulman]{amodei2016concrete}
Dario Amodei, Chris Olah, Jacob Steinhardt, Paul Christiano, and John et~al Schulman.
\newblock Concrete problems in ai safety.
\newblock \emph{arXiv preprint arXiv:1606.06565}, 2016.

\bibitem[Russell(2019)]{russell2019human}
Stuart Russell.
\newblock \emph{Human compatible: {AI} and the problem of control}.
\newblock Viking, 2019.

\bibitem[Russell(2021)]{russell2021artificial}
Stuart Russell.
\newblock Artificial intelligence and the problem of control.
\newblock \emph{Perspectives on Digital Humanism}, pages 19--24, 2021.

\bibitem[Carroll et~al.(2023)Carroll, Chan, Ashton, and Krueger]{carroll2023characterizing}
Micah Carroll, Alan Chan, Henry Ashton, and David Krueger.
\newblock Characterizing manipulation from ai systems.
\newblock In \emph{Proceedings of the 3rd ACM Conference on Equity and Access in Algorithms, Mechanisms, and Optimization}, pages 1--13, 2023.

\bibitem[Lazar and Nelson(2023)]{lazar2023ai}
Seth Lazar and Alondra Nelson.
\newblock Ai safety on whose terms?
\newblock \emph{Science}, 381\penalty0 (6654):\penalty0 138--138, 2023.

\end{thebibliography}

\bibliographystyle{unsrtnat}

\end{document}